\documentclass[11pt,a4paper]{article}
\usepackage{amssymb}
\usepackage{amsmath}\usepackage{graphicx}  
\numberwithin{equation}{section}
\usepackage{footnote}
\makesavenoteenv{tabular}

\newcommand{\id}[1]{\ensuremath{\mathrm{id}}}

\newcommand{\half}{\mbox{\footnotesize $\frac{1}{2}$}}
\newcommand{\hi}[1]{\emph{\textbf{#1}}}
\newcommand{\qm}{quantum mechanics}
\newcommand{\er}{\eqref}
\renewcommand{\L}{\label} 
\newcommand{\beq}{\begin{equation}}
\newcommand{\eeq}{\end{equation}} 
\newcommand{\bea}{\begin{eqnarray}}
\newcommand{\eea}{\end{eqnarray}} \newcommand{\nn}{\nonumber}

\newcommand{\ul}{\underline}

 \newcommand{\til}{\tilde}
\newcommand{\raw}{\rightarrow}

 \newcommand{\Raw}{\Rightarrow}

\newcommand{\lraw}{\leftrightarrow}

\newcommand{\ot}{\otimes} 
\newcommand{\la}{\langle} \newcommand{\ra}{\rangle}

\newcommand{\x}{\times}

\newcommand{\ca}{C*-algebra}

\newcommand{\Hs}{Hilbert space}

\newcommand{\al}{\alpha} 
 
 \newcommand{\Dl}{\Delta}

\newcommand{\lm}{\lambda} \newcommand{\Lm}{\Lambda}
\newcommand{\rh}{\rho} \newcommand{\sg}{\sigma}
 \newcommand{\ta}{\tau} 
 \newcommand{\phv}{\varphi}
\newcommand{\ch}{\ch} \newcommand{\ps}{\psi} 
\newcommand{\om}{\omega} \newcommand{\Om}{\Omega}

\newcommand{\inv}{^{-1}}

\newcommand{\Tr}{\mbox{\rm Tr}\,}

 \newcommand{\CS}{{\mathcal S}}
 \newcommand{\CP}{{\mathcal P}}

\renewcommand{\L}{\label}

\newcommand{\C}{{\mathbb C}} 
\newcommand{\N}{{\mathbb N}} \newcommand{\R}{{\mathbb R}}
 
  \makeatletter
 
\makeatletter
\def\moverlay{\mathpalette\mov@rlay}
\def\mov@rlay#1#2{\leavevmode\vtop{%
   \baselineskip\z@skip \lineskiplimit-\maxdimen
   \ialign{\hfil$\m@th#1##$\hfil\cr#2\crcr}}}
\newcommand{\charfusion}[3][\mathord]{
    #1{\ifx#1\mathop\vphantom{#2}\fi
        \mathpalette\mov@rlay{#2\cr#3}
      }
    \ifx#1\mathop\expandafter\displaylimits\fi}
\makeatother

\newtheorem{definition}{definition}[section]

\newtheorem{theorem}[definition]{Theorem}

\newtheorem{corollary}[definition]{Corollary}
\topmargin = - 1 cm \textheight = 23 cm \textwidth = 15 cm
\begin{document} 
\pagenumbering{arabic} \setlength{\unitlength}{1cm}\cleardoublepage
\date\nodate
\begin{center}
\begin{huge}
{\bf Randomness? What randomness?}
\end{huge}
\bigskip

\bigskip

\renewcommand{\thefootnote}{\alph{footnote}}
\begin{Large}
 Klaas Landsman\footnote{Department of Mathematics, 
Institute for Mathematics, Astrophysics, and Particle Physics (IMAPP), Faculty of Science, Radboud University, Nijmegen, The Netherlands, and Dutch Institute for Emergent Phenomena (DIEP), \texttt{www.d-iep.org}.
Email:
\texttt{landsman@math.ru.nl}.}
\end{Large}
\bigskip

\begin{large}
\emph{Dedicated to Gerard 't Hooft, on the 20th anniversary of his Nobel Prize}\footnote{This paper is an extended version of my talk on July 11th, 2019 at the conference \emph{From Weak Force to Black Hole Thermodynamics and Beyond} in Utrecht in honour of Gerard 't Hooft. I am indebted to him,  Jeremy Butterfield,  Sean Gryb, Ronnie Hermens, Ted Jacobson, Jonas Kamminga, Bryan Roberts, Carlo Rovelli, Bas Terwijn,  Jos Uffink,  Noson Yanofsky, and especially Guido Bacciagaluppi and Cristian Calude for (often contradictory) comments and questions, which have led to numerous  insights and improvements.  }
\end{large}
\end{center}
\bigskip

 \begin{abstract} 
\noindent 
This is a review of the issue of randomness in quantum mechanics, with special emphasis on its ambiguity; for example, randomness has different antipodal relationships to  determinism, computability, and compressibility. Following a (Wittgensteinian) philosophical discussion of randomness in general, I 
 argue that deterministic interpretations of quantum mechanics (like Bohmian mechanics or 't Hooft's Cellular Automaton interpretation) are strictly speaking incompatible with the Born rule.  
I also stress the role of outliers, i.e.\ measurement outcomes that are \emph{not} 1-random. Although these occur with low (or even zero) probability, their very existence implies  that the no-signaling principle used in proofs of randomness of outcomes of quantum-mechanical measurements (and of the safety of quantum cryptography) should be reinterpreted statistically, like the second law of thermodynamics. In three appendices I 
discuss the Born rule and its status in both single and repeated experiments,  review the notion of 1-randomness (or algorithmic randomness) that in various guises was investigated by Solomonoff, Kolmogorov, Chaitin, Martin-L\"{o}f, Schnorr, and others, and treat Bell's (1964) Theorem and the Free Will Theorem with their implications for randomness.  
\end{abstract}
\bigskip
\tableofcontents

\thispagestyle{empty}
\renewcommand{\thefootnote}{\arabic{footnote}}
\newpage \setcounter{footnote}{0}
\section{Introduction}
\begin{quote}\begin{small}
Quantum mechanics commands much respect. But an inner voice tells me that it is not the real McCoy. The theory delivers a lot, but it hardly brings us closer to God's secret. Anyway, I'm sure \emph{he} does not play dice. (Einstein to Born, 1926).\footnote{The German original is: `Die Quantenmechanik ist sehr achtung-gebietend. Aber eine innere Stimme sagt mir, da\ss\ das doch nicht der wahre Jakob ist. Die Theorie liefert viel, aber dem Geheimnis des Alten bring sie uns kaum n\"{a}her. Jedenfalls bin ich \"{u}berzeugt, da\ss\ \emph{der} nicht w\"{u}rfelt.' (Translation by the author.) The source is Einstein's letter to Max Born from December 4, 1926, see Einstein \& Born (2005), p.\ 154. Note the italics in \emph{der}, or \emph{he} in English: Einstein surely means that it is not \emph{God} (or ``the Father") but the \emph{physicists} who play dice. Since some authors claim the opposite, it may be worth emphasizing that this complaint against the indeterminism of \qm\ as expressed in Born (1926), to which Einstein replies here (see also \S\ref{RQM} below for the specific passage that must have upset Einstein), forms the sole contents of this letter; Einstein's objections to the non-locality  of the theory only emerged in the 1930s.
Even his biographer Pais (1982, p.\ 440) confuses the issue by misattributing the `God does not play dice' quotation to a letter by Einstein to Lanczos from as late as March 21, 1942, in which at one stroke he complains
that `It seems hard to look into God's cards. But I cannot for a moment believe that He plays dice and makes use of ``telepathic" means (as the current quantum theory alleges He does).'}

In our scientific expectations we have grown antipodes. You believe in God playing dice and I in perfect laws in the world of things existing as real objects, which I try to grasp in a wildly speculative way. (Einstein to Born, 1944).\footnote{Letter dated November 7, 1944. Strangely, this letter is not contained in the  \emph{Einstein--Born Briefwechsel 1916--1955} cited in the previous footnote; the source is Born (1949, p.\ 176) and the translation is his.}
\end{small}
\end{quote}
Einstein's idea of `perfect laws' that should in particular be deterministic is also central to 't Hooft's view of physics, as exemplified by his intriguing \emph{Cellular Automaton Interpretation of Quantum Mechanics} ('t Hooft, 2016). One aim of this paper is to provide arguments against this view,\footnote{\label{Emfn} What I will not argue for here is the real reason I do not believe in perfect laws, namely the idea of \emph{Emergence}, according to which there are no fundamental laws, let alone perfect ones: every (alleged) law originates in some lower substratum, which itself is subject to laws originating in yet another realm. As I like to say: `Nothing in science makes sense except in the light of emergence' (free after Theodosius Dobzhansky, who famously wrote that `nothing in biology makes sense except in the light of evolution').} but even if these turn out to be unsuccessful I hope to contribute to the debate about the issue of determinism versus randomness  by providing a broad view of the latter.\footnote{See Bricmont \emph{et al} (2001),
Bub \& Bub (2018), Cassirer (1936),  Frigg (2016), Loewer (2001), Nath Bera \emph{et al} (2017),  Svozil (1993, 2018), and Vaidman (2014)
for  other perspectives on randomness in physics. }
My point in \S\ref{RFR} is that randomness is a Wittgensteinian family resemblance (Sluga, 2006; Baker \& Hacker, 2005, Ch.\ XI), \emph{but a special one  that is always defined through its antipode}, which may change according to the specific use of the term. 

The antipode defining which particular notion of randomness is meant may vary even within \qm, and here two main candidates arise (\S\ref{RQM}): one is \emph{determinism}, as emphatically meant by Born (1926) and most others who claim that randomness is somehow `fundamental' in quantum theory, but the other is \emph{compressibility} or any of the other equivalent notions defining what is called 1-randomness in mathematics as its antipode (see Appendix \ref{AR} for an explanation of this). The interplay between these different notions of randomness is the topic of \S\S\ref{PP}-\ref{CAC}.
In  \S\ref{CAC} I argue that one cannot eat one's cake and have it, in the sense of having a deterministic hidden variable theory underneath \qm\ that is strictly compatible with the Born rule. I also propose, more wildly, that Einstein's prohibition of superluminal signaling should be demoted from an absolute to a statistical law, much as the second law of thermodynamics. 
My analysis  relies on some mathematical background presented  in (independent) Appendices \ref{BR}--\ref{FWT} on the Born rule, Algorithmic (or 1-) randomness, and the Bell and Free Will Theorems.
\section{Randomness as a family resemblance}\label{RFR}
\begin{quote}\begin{small}
The idea that in order to get clear about the meaning of a general term one had to find the common element in all its applications has shackled philosophical investigation; for it has not only led to no result, but also made the philosopher dismiss as irrelevant the concrete cases, which alone could have helped him to understand the usage of the general term.  (Wittgenstein, \emph{Blue Book},  \S\S 19--20). \end{small}
\end{quote}
\begin{quote}\begin{small}
I'm saying that these phenomena have no one thing in common in virtue of which we use the same word for all -- but there are many kinds of \emph{affinity} between them.  (\ldots) we see a complicated network of similarities overlapping and criss-crossing: 
similarities in the large and in the small. I can think of no better expression to characterize these similarities than ``family resemblances"; for the various resemblances between members of a family -- build, features, colour of eyes, gait, temperament, and so on and so forth -- overlap and criss-cross in the same way.    (Wittgenstein, \emph{Philosophical Investigations}, \S\S 65--67). \end{small}
\end{quote}
Though he did not mention it himself, randomness seems a prime example of a phenomenon Wittgenstein would call a `family resemblance'.\footnote{This may be worth emphasizing, since even first-rate philosophers like Eagle (2005) still try to nail it down, ironically citing other philosophers who also did precisely that, but allegedly in the `wrong' way!} Independently, as 
 noted by historians L\"{u}thy \& Palmerino (2016) on the basis of examples from antiquity and medieval thought,\footnote{See Vogt (2011) for a comprehensive survey of the historical usage of randomness and chance etc., including references to original sources. See also L\"{u}thy \& Palmerino (2016) for a brief summary. }  the various different meanings of randomness (or chance) can all be identified by their antipode. Combining these ideas, I submit that randomness is not just any old Wittgensteinian family resemblance, but a  special one that is \emph{always} defined \emph{negatively}:
 \begin{itemize}
\item To begin with, in Aristotle's famous example of a man who goes to the market and  walks into his debtor, the randomness of the encounter derives from the fact that the man did \emph{not} go the the market in order to meet his debtor (but instead went there to buy food). Similarly for the man who digs a hole in his garden to plant a tree and finds a treasure. Even the birth of female babies (and certain other `chance substances' for which he literally uses the Greek word for `monsters') was identified by The Philosopher  as a \emph{failure} of purpose in Nature. Thus what actually happened in all these examples was accidental because (as we would say it) it was \emph{not} intended, or, in Aristotelian parlance, because there was \emph{no} final cause. By the same token, Aristotle found the atomistic cosmos of Democritus ``random" because it was purposeless,  ridiculing him for making the cosmic order a product of chance. 
\item  In contrast,  half a century later Epicurus found the atomic world not random at all and introduced randomness through the \emph{swerve}, immortalized by Lucretius:
\begin{quote}\begin{small}
When the atoms are traveling straight down through empty space by their own weight, at quite indeterminate times and places they swerve ever so little from their course, just so much that you can call it a change of direction. If it were not for this swerve, everything would fall downwards like raindrops through the abyss of space. No collision would take place and no impact of atom on atom would be created. Thus nature would never have created anything. (Lucretius, \emph{De Rerum Natura}, Book II).\footnote{ Lucretius (1951), p.\ 66.  See also Greenblatt (2011) for the thrilling rediscovery of \emph{De Rerum Natura}.} \end{small}
\end{quote} 
This was, so to speak, the first complaint against determinism (the goal of Epicurus/Lucretius was to make room for free will in a world otherwise seen as effectively dead because of the everlasting sequence of cause and effect), and indeed, in the context of our analysis, the key point is that the swerve is random because it is \emph{indeterminate}, or because the atoms \emph{depart} from their natural straight course. 
\item Neither of these classical meanings is at all  identical with the dominant usage from medieval times to the early 20th century, which was exemplified by Spinoza, who  claimed that not only miracles, but also circumstances that have concurred by chance are reducible to \emph{ignorance of the true causes of phenomena}, for which ultimately the will of God (`the sanctuary of ignorance') is invoked as a placeholder.\footnote{See \emph{Ethics}, Part I, Appendix.} 
Thus Spinozist  randomness lies in the \emph{absence} of full knowledge of the entire causal chain of events.
\item In the Leibniz--Clarke correspondence,\footnote{See also  L\"{u}thy \& Palmerino (2016), \S 2.7 for part of the following analysis. The Leibniz--Clarke correspondence is available in many editions, such as Vailati (1997), and the online edition Bennett (2017).}  the latter, speaking for Newton, meant \emph{involuntariness} by randomness. Against Leibniz,
Clarke (and Newton) denied at least God could be  limited like that.
 Leibniz, on the other hand, in some sense ahead of his time (yet in another following  Epicurus/Lucretius),  used the word `random' to designate the \emph{absence} of a determining cause--a possibility which he (unlike Epicurus/Lucretius) denied on the basis of his principle of sufficient reason.\footnote{Hacking (1990, Ch.\ 2) calls this the \emph{doctrine of necessity} and shows it pervaded early modern thought.} This is clear from an interesting passage which is not widely known and predates Laplace:
 \begin{quote}\begin{small}
One sees then that everything proceeds mathematically - that is, infallibly - in the whole wide world, so that if someone could have sufficient insight into the inner parts of things, and in addition could have remembrance and intelligence enough to consider all the circumstances and to take them into account, then he would be a prophet and would see the future in the present as in a mirror. (Leibniz).\footnote{The undated German original is quoted by Cassirer (1936), pp.\ 19--20: 
`Hieraus sieht man nun, das alles mathematisch, d.i. uhnfehlbar zugehe in der ganzen weiten Welt, so gar, dass wenn einer eine genugsame Insicht in die inneren Teile der Dinge haben k\"{o}nnte, und dabei Ged\"{a}chtnis und Verstand genug h\"{a}tte, um alle Umst\"{a}nde vor zu nehmen und in Rechnung zu bringen, w\"{u}rde er ein Prophet sein, und in dem Gegenw\"{a}rtigen das Zuk\"{u}nftige sehen, gleichsam als in einem Spiegel.' English translation by the author. 
} 
\end{small}\end{quote}
\item Arbuthnot, a younger contemporary and follower of Newton, may have been one of the first authors to explicitly address the role of randomness in the deterministic setting of Newtonian physics. In the Preface to his  translation of Huygens's path-breaking book  \emph{De Ratiociniis in Ludo Aleae} on probability theory, he wrote:
\begin{quote}
\begin{small}
It is impossible for a Die, with such determin'd force and direction, not to fall on such determin'd side, only I don't know the force and direction which makes it fall on such determin'd side, and therefore I call it Chance, which is nothing but the want of Art. (Arbuthnot, 1692).\footnote{The Latin original \emph{De Ratiociniis in Ludo Aleae} is from 1657 and Arbuthnot's English translation \emph{On the Laws of Chance} appeared in 1692. The quotation is taken from Diaconis \& Skyrms (2018), p.\ 9. }
\end{small}
\end{quote}
And similarly, but highlighting the alleged negativity of the  concept even more:
\begin{quote}\begin{small}
\emph{Chance}, in atheistical writings or discourse, is a sound utterly insignificant: It imports no determination to any \emph{mode of Existence}; nor in deed to \emph{Existence} itself, more than to \emph{non existence}; it can neither be defined nor understood; nor can any Proposition concerning it be either affirmed or denied, excepting this one, ``That it is a mere word." (De Moivre, 1718).\footnote{Quoted  by Hacking (1990, p.\ 13) from De Moivre's \emph{Doctrine of Chance}, originally written in English.}\end{small}
\end{quote}
So this is entirely in the medieval spirit, where 
ignorance--this time relative to Newton's physics as the ticket to full knowledge--is seen as the origin of randomness.
\item A century later, and like Arbuthnot and De Moivre again in a book on probability theory (\emph{Essai philosophique sur les probabilit\'{e}s}, from 1814), 
Laplace portrayed his  demon  to make the point that randomness arises in the \emph{absence} of such an intellect:
\begin{quote}\begin{small}
An intelligence which could comprehend
all the forces that set nature in motion, and all positions of all items of which nature is composed--an intelligence sufficiently vast to submit these data to analysis--it
would embrace in the same formula the movements of the greatest bodies
in the universe and those of the lightest atom; for it, nothing would be
uncertain and the future, as well as the past, would be present to its eyes. (Laplace, 1814).\footnote{The translation is from Laplace (1902), p.\ 4. See van Strien (2014) for history and  analysis.}  \end{small}
\end{quote}
Note that Leibniz' prophet appeals to the logical structure of the universe that makes it deterministic, whereas Laplace's intelligence  knows (Newtonian) physics.\footnote{However, van Strien (2014) argues that Laplace also falls back on Leibniz (and gets the physics wrong by not mentioning the momenta that the intelligence should know, too, besides the forces and positions).} In any case, it is important to note that Laplacian randomness is defined \emph{within a deterministic world},\footnote{Famously: `The world $W$ is Laplacian deterministic just in case for any physically possible world $W'$, if $W$ and $W'$  agree at any time, then they agree at all time.' (Earman, 1986, p. 13).} so that its 
antipode is not indeterminism but full knowledge (and computing power, etc.). Indeed, less well known than the above quotation is the following one, from the very same source:
\begin{quote}\begin{small}
All events, even those which on account of their insignificance do not seem to follow the great laws of nature, are a result of it just as necessarily as the revolutions of the sun.  (Laplace, 1814).\footnote{Quoted by Hacking (1990), p.\ 11.}
 \end{small}\end{quote}
\item The ignorance interpretation of randomness and chance still lurks behind  the probabilities introduced in the 19th century in statistical mechanics, which in my view were therefore  wholeheartedly construed in the medieval and early modern sprit.\footnote{ It is often maintained that these probabilities are \emph{objective},
 which might cast doubt over the idea that they originate in ignorance. I personally find the distinction between  ``objective"  and ``subjective" chances quite unhelpful in the context of fundamental physics, a left-over from (by now) irrelevant and outdated attempts to define probabilities ``objectively" as relative frequencies (see also Appendix \ref{AR}) as opposed to interpreting them ``subjectively" as credences \`{a} la Ramsey (1931).
 Loewer (2001, 2004) claims that the probabilities in statistical mechanics are as objective as those in \qm\ (and hence are objective). Indeed, both are predicated on a  choice of observables,  based on ignoring microscopic and quantum-mechanical degrees of freedom, respectively. \emph{Given} that choice, the probabilities are  objective, but the \emph{choice}  is surely subjective (see also Heisenberg quoted in \S\ref{RQM} below). Instead of a pointless table-tennis game between ``objective" and ``subjective",  a better term would be ``perspectival". For example, I agree with Rovelli (2017) that time's arrow is perspectival (but neither  ``objective" nor ``subjective"). See also Jaynes (1985),  pp.\ 118--120, for similar comments on  terminology in the philosophy of probability.
}
 \item Hacking (1990) explains  how the doctrine of necessity began to erode in the 19th century, largely through the use of statistics in population studies and biology.\footnote{Though also inspired by the eventual rise of \qm, Kern (2004) gives a completely different history of causality and uncertainty in the 19th and 20th centuries, tracking the changing explanatory roles of these factors in the the study of murder as documented by more than a hundred novels.} In this respect, the century
to some extent  culminated in the following words of Peirce:
 \begin{quote}\begin{small}
I believe I have thus subjected to fair examination all the important reasons for adhering to the theory of universal necessity, and shown their nullity.  (\ldots) If my argument remains unrefuted,
it will be time, I think, to doubt the absolute truth of the principle of universal law. (Peirce, 1892, p.\ 337).
\end{small}\end{quote}
 This partly paved the way for the claim of irreducible randomness in \qm, although the influence of population studies and biology on intrinsic developments in physics should perhaps not be overestimated. However, the insight that probability and statistics gave rise to their own laws (as opposed to the fear that randomness is pretty much the same as lawlessness), which partly dated back to the previous two centuries (Hacking, 2006),
 surely made quantum \emph{theory} possible.
\item The randomness of variations in heritable traits that--almost simultaneously with the rise of statistical physics--were introduced in Darwin's theory of evolution by natural selection meant something completely different from Laplace etc., best expressed by the renowned geneticist Theodosius Dobzhansky a century later (cf.\ Merlin, 2010):
\begin{quote}\begin{small}
 Mutations are random changes because they occur independently of whether they are beneficial or harmful.
 (Dobzhansky \emph{et al}, 1977, p.\ 66).
\end{small}\end{quote}
 Indeed, both historically and actually, the antipode to Darwin's randomness of variations  is Lamarck's goal-orientedness thereof,\footnote{ As such, Spinoza's philosophical analysis, modern physics, and (evolutionary) biology all contributed to the downfall of the Aristotelian (and subsequently Christian)  final causes my list started from. } intended to strengthen the species (like the proverbial sons of the  blacksmith who according to Lamarck inherit his strong muscles). In particular, it does not matter if the variations are of known or unknown origin, or fall under a deterministic or an indeterministic kind of physics.
 \item Continuing our detour into biology,  the well-known Hardy--Weinberg equilibrium law in population genetics, which gives  the relative frequencies of alleles and genotypes in a given (infinite) population, is based on the assumption of \emph{random mating}. This means that mating takes place between pairs of individuals who have \emph{not} selected each other on the basis of their genetic traits (i.e.\ there is \emph{no} sexual selection).
 \item Eagle (2005, p.\ 775--776) proposes that `randomness is maximal \emph{un}predictability' (which agrees with criterion 3 at the end of this section), and argues that this is equivalent to 
a random event being `probabilistically \emph{in}dependent of the current and past states of the system, given the probabilities
supported by the theory.' 
 \item Most people, especially those without scientific training,\footnote{Though Stephen Hawking was not adverse to the fact that he was born on January 8th, 1942, exactly 300 years after the death of Galileo Galilei, and, had he been able to note it, would undoubtedly have rejoiced in the equally remarkable fact that he died in 2018 on the birthday of Albert Einstein
(March 14).}
  asked to mention a random event they have encountered, typically mention what is called a \emph{coincidence}. 
  This notion goes back at least to Aristotle, but a modern definition is the following:
  \begin{quote}\begin{small}
  A coincidence is a surprising concurrence of events, perceived as meaningfully related, with no apparent causal connection. (Diaconis \& Mosteller, 1989,  p.\,853).\end{small}\end{quote}
The subjective nature of this definition (i.e.\ `surprising', `perceived', `meaningful', and `apparent') may  repel the scientist, but ironically, the aim of this definition is to debunk coincidences. Moreover, there is a striking similarity between analyzing
 coincidences in daily life and coincidences in the physical setting of bipartite correlation experiments \`{a} la EPR--Bohm--Bell: to this end, let me briefly recall how most if not all everyday coincidences can be nullified on the basis of the above definition:\footnote{For details see Landsman (2018), so far available in Dutch only. In practice it usually does not help to argue that the events in question were not `meaningfully related', since at that point  semi-intellectual opponents will invoke ``synchronicities" (Jung, 1952) and the like, and leave the domain of science (as Jung himself squarely admitted). 
Even if one goes along with that, discussions tend to become circular.
}
\begin{enumerate}
\item \emph{Against first appearances there \emph{was} a causal connection, either through a common cause or through direct causation.} This often works in daily life, and also in Bohmian mechanics, where direct (superluminal) causation is taken to be the ``explanation" of the correlations  in the experiments just referred to. However, if superluminal causation is banned, then one's hand is empty because  one interpretation of Bell's Theorem (cf.\ Appendix \ref{FWT}) excludes common causes (van Fraassen, 1991), and hence both kinds of causation are out!  See also  \S\ref{RQM}. 
\item \emph{The concurrence of events was not at all as surprising as initially thought.} This argument is either based on the inability of most people to estimate probabilities correctly (as in the well-known  birthday problem), or, if the events were really unlikely, on
  what Diaconis and Mosteller call \emph{the law of truly large numbers}:\footnote{In a slightly different phrasing this ``law"  is also called  \emph{The Improbability Principle} (Hand, 2015).}
  \begin{quote}\begin{small}
With a large enough sample, any outrageous thing is likely to happen.  (Diaconis \& Mosteller, 1989, p.\ 859).
   \end{small}\end{quote}
\end{enumerate}
  Neither of this helps in  ERR--Bohm--Bell, though, leaving one's hand \emph{truly} empty.
\item The \emph{choice sequences} introduced by Brouwer in his intuitionistic construction of the real numbers, and then especially the ``lawless" ones (which Brouwer never defined precisely) are often associated with notions of ``freedom" and ``randomness":
\begin{quote}\begin{small}
A choice sequence in Brouwer's sense is a sequence of natural numbers (to keep it simple), which is not a priori given by a law or recipe for the elements, but which is created step by step (\ldots); the process will go on indefinitely, but it is never finished.  (\ldots) Informally, we think of a lawless sequence of natural numbers as a process of choosing values in $\N$ (\ldots) under the a priori restriction that at any stage of the construction never more that an initial segment has been determined and that \emph{no} restrictions have been imposed on future choices, [and that] there is a commitment to determine more and more values (so the sequence is infinite). (\ldots)
\emph{A lawless sequence may be compared to the sequence of casts of a die}. There too, at any given stage in the generation of a sequence never more than an initial segment is known. (Troelstra, 1996,  {italics added}).\footnote{See also Troelstra (1977) and references therein to the original literature.}
\end{small}
\end{quote}
 Indeed,  von Mises (1919) mentioned choice sequences as an inspiration for his idea of a \emph{Kollektiv}  (van Lambalgen, 1996), which in turn paved the way for the theory of algorithmic randomness to be discussed as the next and final example of this section.
 
Nonetheless, a more precise analysis (Moschovakis, 1996, 2016) concludes as follows:
\begin{quote}\begin{small}
\emph{Lawless} and \emph{random} are orthogonal concepts. A random sequence of natural numbers should possess certain definable regularity properties (e.g.\ the percentage of even numbers in its $n$th segment should apporach 50 as $n$ increases),\footnote{See for example Theorem \ref{PRS} below.} while a lawless sequence should possess none. Any regularity property definable in $\mathcal{L}$ by a restricted formula can be defeated by a suitable lawlike predictor.
 (Moschovakis, 2016, pp.\ 105--106, italics in original, footnote added).
 \end{small}
\end{quote}
 A further \emph{conceptual} mismatch between lawless choice sequences  and random sequences of the kind studied in 
 probability theory is that any randomness of the  former seems to lie on the side of what is called \emph{process randomness}, whereas the latter is concerned with \emph{product randomness}.\footnote{See e.g.\ Eagle (2005) for this terminology. } In a lawless choice sequence it is its  \emph{creation process} that is lawless, whereas no \emph{finished} sequence (i.e.\ the \emph{outcome}) can be totally lawless by  \emph{Baudet's Conjecture}, proved by van der Waerden (1921), which implies that \emph{every}  binary sequence $x$ satisfies certain arithmetic laws.

  \item Finally, serving our aim to compare physical and mathematical notions of randomness, I preview the three equivalent definitions of 1-randomness (see Appendix \ref{AR} and references therein  for details) and confirm that also they  fit into our general picture of randomness being defined by some antipode. What will be remarkable is that the three apparently  different notions of randomness to be discusses now, which at first sight are as good a family resemblance as any, actually turn out to \emph{coincide}. The objects whose randomness is going to be discussed are  binary strings, and our discussion here is so superficial that I will not even distinguish between finite and infinite ones; see Appendix \ref{AR} for the kind of precision that does enable one to do so. 
  \begin{enumerate}
\item A string $x$ is 1-random if its shortest description is $x$ itself, i.e.,  there exists \emph{no} lossless compression of $x$ (in the sense of a computer program that outputs $x$ and whose length is shorter than the length of $x$): thus $x$ is \emph{incompressible}.
\item A string $x$ is 1-random if it \emph{fails} all tests for patterns (in a computable  class).
\item A string $x$ is 1-random if there exists \emph{no} successful (computable)  gambling strategy on the digits of $x$; 
roughly speaking, these digits are \emph{unpredictable}.
\end{enumerate}
  \end{itemize}
    \section{Randomness in quantum mechanics}\label{RQM}
    Moving towards the main goal of the paper, I now continue our list of  examples (i.e.\ of the principle that randomness is a family resemblance whose different meanings are always defined negatively through their antipodes) in the context of quantum mechanics, which is rich enough by itself to provide its own family of different meanings of 
 randomness (all duly defined negatively), although these may eventually be traceable to the above cases.\footnote{ \label{fnpp}It will be clear from the way I discuss  \qm\ that this paper will just be concerned with \emph{process randomness}, as opposed to \emph{product randomness}. There is certainly a distinction between the two, but the former  drags  us into the quagmire of the measurement problem (cf.\ Landsman, 2017, Chapter 11).}
 Already the very first (scholarly) exposition of the issue of randomness in quantum mechanics by Born made many of the major points that are still relevant today:
 \begin{quote}\begin{small}
Thus Schr\"{o}dinger's quantum mechanics gives  a  very definite answer to the 
question of  the  outcome  of  a  collision;  however, this does not involve any  causal  relationship.   One obtains 
\emph{no} answer  to  the  question  ``what  is  the  state  after  the  collision,"
but only to the question ``how probable is a specific outcome
 of the collision" (in which the quantum-mechanical law of [conservation of] energy must of course be satisfied). 
This raises the entire problem of determinism.  From the standpoint of our quantum 
mechanics, there is no quantity that could causally establish the outcome of a collision  
in each individual case;  however,  so far we are not aware of any experimental clue to the effect 
that there are  internal  properties  of  atoms  that enforce some particular outcome.  Should we hope to discover such properties that determine individual outcomes later
(perhaps phases of the internal  atomic  motions)?  Or should  we  believe  that  the agreement  between  theory  and  experiment concerning our  inability  to give  conditions for 
a causal course of events is some pre-established harmony that is based on the non-existence of
such conditions? I myself tend to relinquish determinism in the atomic world. But this is a philosophical question, for which physical arguments alone are not decisive. \\ (Born, 1926, p.\ 866).\footnote{Translation by the author. The reference is to the German original.}
\end{small}\end{quote}
   Given the fact that Born was the first to discuss such things in the open literature, it is remarkable how perceptive his words are: he marks the opposition of randomness to determinism, recognizes the possibility of hidden variables (with negative advice though), and understands that the issue  is not just a technical one. \emph{Bravo! }Having said this, in line with the previous section our aim is, of course, to confront the antipode of determinism with other possible antipodes to randomness as it is featured by \qm. 
   
The introduction of fundamental probabilities in quantum theory is delicate in many ways, among which is the fact that the Schr\"{o}dinger equation is even more deterministic than Newton's laws.\footnote{\label{STE} In the sense that the solution is not only uniquely determined by the initial state, but, by Stone's Theorem,  even exists for all $t\in\R$; in other words, incomplete motion is impossible, cf.\ Earman (2009).} Hence what is meant  is randomness of \emph{measurements outcomes};
since it is not our aim (here) to solve the measurement problem--for which see Landsman (2017), Chapter 11--I simply assume that \emph{i)} measurement is  a well-defined laboratory practice, and \emph{ii)}  measurements have outcomes. In all that follows, I also accept the statistical predictions of \qm\ for these outcomes (which are based on the Born rule reviewed in Appendix \ref{BR}).
  Even so, the claim of  \emph{irreducibility} of randomness, which is typical for all versions of the  Copenhagen Interpretation (and for any mainstream view held by physicists) is almost  incomprehensible, since one of the pillars of this interpretation is  Bohr's \emph{doctrine of classical concepts}, according to which the apparatus must be described classically;  randomness of measurement outcomes is then seen as a consequence of the very definition of a measurement. But this 
  suggests that randomness should be \emph{reducible} to ignorance about the quantum-mechanical degrees of freedom of the apparatus:
 \begin{quote}\begin{small}  these uncertainties (\ldots) are simply a consequence of the fact that we describe the experiment in terms of classical physics.
(Heisenberg, 1958, p.\ 53).
\end{small}\end{quote} 
Ironically, Bell's Theorem(s), which arose in opposition to the  the Copenhagen Interpretation (Cushing, 1994), did not only prove Einstein (as the leading opponent of this interpretation) \emph{wrong} on the issue that arguably mattered most to him (namely locality in the sense defined later in a precise way by Bell), but also proved Bohr and Heisenberg \emph{right} on the \emph{irreducibility} of randomness, at least if we grant them randomness \emph{per se}. Indeed, suppose we equate  \emph{reducibility} of randomness
with the existence of a ``Laplacian" deterministic hidden variable theory (i.e.\ use the antipode of determinism),
and assume, as the Copenhagenists would be pleased to,  the conjunction of the following properties:
\begin{enumerate}
\item  The  \emph{Born rule} and the ensuing statistical predictions of \qm;
\item \emph{(Hidden) Locality}, i.e.\ the impossibility of active superluminal communication or causation if one knows the state $\lm$  of the underlying deterministic theory;\footnote{This is often called \emph{Parameter Independence}, as in e.g.\ the standard textbook by Bub (1997).}
\item  \emph{Freedom} (or \emph{free choice}), that is,  the independence of the choice of measurement settings from the state of the system one measures using these settings, in a broad sense of `state' that includes the prepared state as well as the ``hidden'' state $\lm$.\footnote{This assumption even makes sense in a super-deterministic theory, where the settings are not free.}
\end{enumerate}
Bell's (1964) Theorem then implies (robustly) that  a deterministic hidden variable satisfying these assumptions theory cannot exist, as does the so-called Free Will Theorem (which relies on a non-probabilistic but non-robust special case of the Born rule implying perfect correlation, and also  on the Kochen--Specker Theorem, which restricts its validity to quantum systems with three levels or more).\footnote{See Cator \& Landsman (2014), Landsman (2017), Chapter 6, or Appendix \ref{FWT} below for a unified view of both Bell's Theorem  (from 1964)  and the Free Will Theorem (named as such by Conway and Kochen).
} Thus the Laplacian interpretation of randomness does not apply to \qm\ 
(granting assumptions 1--3), which warrants the Copenhagen claim of \emph{irreducible} or \emph{non-Laplacian} or \emph{Leibnizian} randomness.
  
   Viable deterministic hidden variable theories compatible with the Born rule therefore have to choose between giving up either \emph{Hidden Locality} or  \emph{Freedom} (see also the conclusion of Appendix \ref{FWT}). 
   Given this choice, we may  therefore distinguish between theories that:
  \begin{itemize}
\item  give up  \emph{Hidden Locality}, like Bohmian mechanics;
 \item  give up  \emph{Freedom}, like the cellular automata interpretation of 't Hooft (2016).
 \end{itemize}
 In both cases the statistical predictions of \qm\ are recovered by averaging the hidden variable or state with respect to 
 a probability measure $\mu_{\ps}$ on the space of hidden variables, given some (pure) quantum state $\ps$. The difference is that in Bohmian mechanics the total state (which consists of the hidden configuration plus the ``pilot wave'' $\ps$) determines the measurement outcomes \emph{given the settings}, whereas in  't Hooft's theory 
 the hidden state (see below) all by itself determines the outcomes as well as the settings.
 \begin{itemize}
\item  In Bohmian mechanics the hidden variable is position $q$, and $d\mu_{\ps}=|\psi(q)|^2 dx$ is  the Born probability for outcome $q$ with respect to the expansion $|\psi\ra=\int dq\, \psi(q) |q\ra$.\footnote{Bohmians call this choice of $\mu_{\ps}$ the  \emph{quantum equilibrium condition}. It was first written down by Pauli.}
\item In 't Hooft's theory the hidden state  is identified with a basis vector $|n\ra$ in some Hilbert space $H$ ($n\in\N$), and once again the measure $\mu_{\ps}(n)=|c_{n}|^2$  is given by the Born probability for outcome $n$ with respect to the expansion $|\psi\ra=\sum_n c_n|n\ra$. 
\end{itemize}
In Bohmian mechanics ('t Hooft does not need this!) such averaging  also restores \emph{Surface Locality}, i.e.,   the impossibility of  superluminal communication on the basis of \emph{actual} measurement outcomes (which is a theorem of quantum theory, though a much more delicate one than is usually thought, as I will argue in \S\ref{CAC}), see also  Valentini (2002ab).
 \section{Probabilistic preliminaries}\label{PP}
\begin{quote}\begin{small}
O False and treacherous Probability,\\
Enemy of truth, and friend to wickednesse;\\
With whose bleare eyes Opinion learnes to see,\\
Truth's feeble party here, and barrennesse.\\
When thou hast thus misled Humanity, \\
And lost obedience in the pride of wit, \\
With reason dar'st thou judge the Deity, \\
And in thy flesh make bold to fashion it. \\
Vaine thoght, the word of Power a riddle is, \\
And till the vayles be rent, the flesh newborne, \\
Reveales no wonders of that inward blisse, \\
Which but where faith is, every where findes scorne; \\
Who therfore censures God with fleshly sp'rit, \\
As well in time may wrap up infinite\\
\mbox{} \\
Philip Sidney (1554--1586), C\oe lica, Sonnet CIV.\footnote{The first four lines of this poem are printed on the last page  of Keynes (1921), without any source.}
\end{small}\end{quote}
My aim is to give a critical assessment of the situation described in the previous section. My analysis is based on the interplay between the single-case probability measure $\mu$ on an outcome space $X$, which for the purpose of this paper will  be the Born measure $\mu=\mu_a$ on the spectrum $X=\sg(a)$ of some self-adjoint operator $a$,  and hence is provided by \emph{theory} (see also Appendix \ref{BR}), and the probabilities defined as long-run frequencies for outcome sequences $x=(x_1, x_2, \ldots)$ of the Bernoulli process defined by $(X,\mu)$, which are given by \emph{experiment}. To obtain clean mathematical results, I assume experiments can be repeated infinitely often. This is clearly  an idealization, which is regulated by \emph{Earman's Principle}:
  \begin{quote}\begin{small}
While idealizations are useful and, perhaps, even essential to progress in physics, a sound principle of interpretation would seem to be that no effect  can be counted as  a genuine physical effect if it disappears
when the idealizations are removed.  (Earman, 2004, p.\ 191).
\end{small}\end{quote}
 As shown in Appendix \ref{AR}, finite-size effects amply confirm the picture below, comparable to the way that
 the law(s) of large numbers have finite-size approximants (such as the Chernoff-Hoeffding bound). In particular, zero/unit probability of  infinite sequence comes down to very low/high  probability for the corresponding finite strings.\footnote{One has to  distinguish between \emph{outcomes} with probability zero and \emph{properties} that hold with probability zero. Indeed, \emph{every} single outcome (in the sense of an infinite bitstream produced by a fair quantum coin flip) has probability zero, and hence this property alone cannot distinguish between random outcomes (in whatever sense, e.g.\ 1-random) and non-random outcomes (such as deterministic outcomes, or, more appropriately in our technical setting, computable ones), or indeed between any kind of different outcomes. On the other hand, not being random is a \emph{property} that holds with probability zero, in that the \emph{set} of all outcomes that are not random has probability zero (equivalently, the event consisting of all outcomes that are random happens almost surely, i.e.\ with probability 1). And yet outcomes with this probability zero property exist and may occur (in the finite case, with very small but positive probability).
 See also \S\ref{CAC}.}
 Consequently, Earman's Principle (in contrapositive form) holds  if we use the canonical probability measure $\mu^{\infty}$ on the 
  infinite product space $X^{\N}$ of all infinite sequences $x=(x_1, x_2, \ldots)$, where $x_n\in X$, 
  and $X^{\N}$ are canonically equipped with the cylindrical  $\sg$-algebra $\CS\subset \CP(X^{\N})$.\footnote{\label{sigma}
Here $\CP(Y)$ denotes the power set of $Y$. The $\sg$-algebra $\CS$ is generated by all subsets of the form 
$B= \prod_{n=1}^N A_n\x \prod_{m=N+1}^{\infty} X$,
where  $A_n\subset X $ is Borel measurable and $N<\infty$. The probability measure $\mu$ on $X$ then defines a probability measure $\mu^{\infty}$ on $\CS$, whose value  on  $B$ is defined by
$\mu^{\infty}(B)=\prod_{n=0}^N \mu(A_n)$. See e.g.\ Dudley (1989), especially Theorem 8.2.2, for this and related constructions (due to Kolmogorov). }

To see how one may recover or verify $\mu$ from the long-term frequencies governed by the product measure $\mu^{\infty}$, for any function $f:X\raw\R$ define  $f^{(N)}:X^N\raw\R$ by 
\begin{equation}
f^{(N)}(x_1, \ldots, x_N)=\frac{1}{N}(f(x_1)+\cdots + f(x_N)). \label{fN}
\end{equation}
Then (by the ergodic theorem), for any $f\in C(X)$, almost surely with respect to $\mu^{\infty}$,
\begin{equation}
\lim_{N\raw\infty}f^{(N)}=\la f\ra_{\mu} \equiv \int_{X}d\mu(x)\, f(x)
,\label{nuas3}
\end{equation}
times  the unit function $1_{X^{\N}}$ on  $X^{\N}$. This is  for continuous functions $f$, but
a limit argument extends it to characteristic functions $1_A:X\raw\ul{2}\equiv \{0,1\}$, where $A\subset X$, so that  
\begin{equation}
\lim_{N\raw\infty} 1_A^{(N)}= \mu(A),  \label{nuas5}
\end{equation}
times $1_{X^{\N}}$, again $\mu^{\infty}$-almost surely. If we define the probability of $A$ within some infinite sequence $(x_1, x_2, \ldots)$ as the relative frequency of $A$ (i.e.\ the limit as $N\raw\infty$ of the number of $x_n$ within $(x_1, \ldots, x_N)$ that lie in $A$ divided by $N$),
then \er{nuas5} states that for almost all sequences  in $X^{\N}$ 
 with respect to the infinite product measure $\mu^{\infty}$
 this probability of $A$ equals its Born probability. This is  useful, since the latter is a purely mathematical quantity, whereas the former is experimentally accessible (at least for large $N$).
 
For simplicity (but without real loss of generality, cf.\  footnote \ref{noloss}), in what follows I specialize this setting to a (theoretically) fair coin flip, that is, $X=\ul{2}=\{0,1\}$ and $\mu(0)=\mu(1)=1/2$. Hence $\ul{2}^{\N}$ is the space of infinite binary sequences, equipped with the probability measure $\mu^{\infty}$ induced by $\mu$ (as I shall argue in \S\ref{CAC} below, despite tradition this situation cannot in fact arise classically, at least not in a deterministic theory). We have:
 \begin{theorem}\label{thm:3.1}
 Almost every binary sequence $x\in\ul{2}^{\N}$ is 1-random with respect to $\mu^{\infty}$.
  \end{theorem}
 See e.g.\ Calude (2010), Corollary 6.32. Thus the set $E$ of all sequences that are \emph{not} 1-random has probability zero, i.e.\ $\mu^{\infty}(E)=0$, but this by no means implies that $E$ is ``small" in any other sense: set-theoretically, it is as large as its complement, i.e.\ the set of all 1-random sequences, which makes it bizarre that (barring a few exceptions related to Chaitin's number $\Om$) not a single one of these  1-random sequences can actually be \emph{proved} to be 1-random, cf.\ Appendix \ref{AR}. Theorem \ref{thm:3.1} has further amazing consequences:
  \begin{corollary}\label{CC}
 With respect to $\mu^{\infty}$, almost every infinite outcome sequence $x$ of a  fair coin flip
 is Borel normal,\footnote{This means that any group $x_1 \cdots x_n$ of digits in $x$ occurs with relative frequency equal to $2^{-n}$. }
 incomputable,\footnote{This means that there is no computable function $f:\N\raw\N$ taking values in $\{0,1\}$ that produces $x$, where computability is meant in the old sense of recursion theory, or, equivalently, in the sense of Turing machines or indeed modern computers. One could also say that there is no algorithm for $f$.} 
  and contains any finite string infinitely often.\footnote{This is a strong version of the \emph{infinite monkey (typewriter) theorem}, to the effect that if some monkey randomly hits the keys of a typewriter for a very long time, the probability that it produces a play by Shakespeare (or whatever other canonical text) is positive. See also the ``Boltzmann Brain argument." }
 \end{corollary}
 This follows because any 1-random sequence has these properties \emph{with certainty}, see Calude (2010), \S 6.4. 
 The relevance of Bernoulli processes for quantum theory comes from Theorem \ref{ET} in the next section, whose second option almost by definition yields these processes.
 \section{Critical analysis and claims}\label{CAC}
 The relevance of the material in the previous section comes from the following result.
  \begin{theorem}\label{ET}
  The following procedures for repeated identical measurements  are equivalent (in giving the same possible outcome sequences with the same  probabilities):
 \begin{enumerate}
 \item {Quantum mechanics is  applied to the whole run (with classically recorded outcomes).}
\item {Quantum mechanics is just applied to single experiments (with classically recorded outcomes), upon which classical probability theory takes over to combine these.} 
\end{enumerate}
 \end{theorem}
See Appendix \ref{BR} for the proof, which culminates in eq.\ \er{mula}, showing that the Born probability $\mu_a$ for single outcomes induces the Bernoulli process probability $\mu_a^{\infty}$ on the space $\sg(a)^{\N}$ of infinite outcome sequences. As mentioned before, I specialize to fair quantum coin flips producing 50-50 Bernoulli processes, of which there are examples galore: think of
measuring the third Pauli matrix $\sg_z=\mathrm{diag}(1,-1)$ in a state like $\psi=(1,1)/\sqrt{2}$. In that case,
Theorem \ref{ET} and Corollary \ref{CC} obviously have the following implication:
\begin{corollary}\label{CC2}
With respect to the product measure
 $\mu_a^{\infty}$ coming from the Born measure $\mu_a$, almost every infinite outcome sequence $x$ of a  fair  quantum coin flip is 1-random and therefore Borel normal, incomputable,
  and contains any finite string infinitely often.
 \end{corollary}
The Born rule therefore implies very strong randomness properties of outcome sequences, albeit merely with probability one (i.e.\ almost surely) with respect to $\mu_a^{\infty}$. Moreover, Chaitin's Incompleteness Theorem (see Appendix \ref{AR}) makes it impossible to prove that some given outcome sequence is 1-random \emph{even if it is}!\footnote{See
Abbott \emph{et al} (2019) and Kovalsky \emph{et al} (2018) for empirical tests of approximate 1-randomness.}
Also  in the spirit of the general `family resemblance' philosophy of randomness, this puzzling situation makes it natural to compare  randomness of infinite 
 measurement outcome sequences as defined by:\footnote{\label{mrd} In the literature on quantum cryptography and quantum random number generators (QRNG) one also find notions of (``free") randomness for single qubits, see e.g.\ the reviews Brunner  \emph{et al} (2019),   Herrero-Collantes \& Garcia-Escartin (2017), Ma \emph{et al} (2016), and Pironio (2018).
The two (closely related) definitions used there fall  into our general scheme of antipodality, namely randomness as \emph{unpredictability} defined as
minimizing the probability that an eavesdropper correctly guesses the qubit, and randomness as the absence of correlations between the qubit in question and anything outside its forward lightcone (which is heavily Leibnizian). These are related, since guessing is done through such correlations of the qubit.
 Certification of both cryptographic protocols and QRNGs is defined  in these terms, backed by proofs of indeterminism \`{a} la Bell, so that in all versions non-locality plays a crucial role. See also footnote \ref{fnqrng}.}
\begin{enumerate}
\item  \emph{1-randomness}  (with compressible sequences as its antipode), as suggested by the Born rule and the above analysis of its mathematical implications;\footnote{I am  by no means the first to relate quantum theory to algorithmic randomness: see, for example,  Bendersky \emph{et al} (2014, 2016, 2017),  Calude (2004), Svozil (1993, 2018),  Senno (2017), Yurtsever (2001), and Zurek (1989). The way my work relates to some of these references will become clear  in due course.}
\item \emph{indeterminism}, as suggested by Born himself, and  in his wake also by  Bell's Theorem and the Free Will Theorem  (seen as proofs  of indeterminism under  assumptions 1--3).
\end{enumerate}
To make this comparison precise, we once again need the case distinction between hidden variable theories giving up Hidden Locality like Bohmian mechanics and those giving up Freedom, like 't Hooft's theory.  See Appendix \ref{FWT} for notation.  In the usual  EPR--Bohm--Bell setting (Bub, 1997), let $(\al,\beta)\in O\x O$ be Alice's and Bob's outcomes for given settings $(a,b)\in S\x S$, so that $(\al,a)$ are Alice's and $(\beta,b)$ are Bob's (outcomes, settings).
\begin{itemize}
\item In Bohmian mechanics, 
 the hidden state $q\in Q$ just pertains to the correlated particles undergoing measurement, whilst  the settings $(a,b)$ are supposed to be ``freely chosen" for each measurement (and in particular are independent of $q$).\footnote{This
 is true if one knows the hidden positions exactly at the time of measurement. At earlier times, the pilot wave = quantum state $\psi$ is needed to make predictions, since $\psi$ determines the trajectories. See e.g.\ Barrett (1999), Chapter 5, for a  discussion of measurement in Bohmian mechanics, including EPR--Bohm.}
  The outcome is then given by  $(\al,\beta)=q(a,b)$. So if we number the consecutive runs of the experiment by $n\in\N=\{1,2, \ldots\}$, then
everything is determined by  functions
 \begin{align}
f_1&: \N \raw Q; & f_1(n)&=q_n;\\
f_2&: \N\raw S\x S; & f_2(n)&=(a_n,b_n), \label{Bsettings}
\end{align}
since these also give the outcome sequence by $(\al_n,\beta_n)=q_n(a_n,b_n)=f_1(n)(f_2(n))$.
 \item    In  't Hooft's  theory, the hidden state $x\in X$ of ``the world" determines the settings as well as the outcomes, i.e.\  $(a,b,\al,\beta)=(a(x),b(x),\al(x),\beta(x))$.  In this case,  the entire situation including the outcome sequence is therefore  determined by a function
  \begin{align}
g: \N\raw X; && g(n)= x_n.
\end{align}
\end{itemize}
A key point in the analysis of the functions $f_i$ and $g$ is the requirement that both theories reproduce the statistical predictions of \qm\ given through the Born rule relative to some pure state $\psi$. As already noted, this is achieved by requiring that $q$ is averaged with respect to some probability measure $\mu_{\psi}$ on $Q$, and likewise $x$ is  averaged with respect to some probability measure $\mu_{\psi}'$ on $X$. If the experimental run is to respect this averaging,  then in Bohmian mechanics
 the map $f_1$  must be ``typical" for the Born-like measure $\mu_{\psi}$ 
(cf.\  D\"{u}rr, Goldstein, \& Zanghi, 1992; Callender, 2007;  Norsen, 2018; Valentini, 2019);\footnote{Appendix \ref{AR} contains a precise
definition of typicality, see in particular point 3 below eq.\ \er{50}.}
 see below for the special problems posed also by $f_2$.
 Analogously to the previous discussion of the Born measure itself and its sampling, any sampling of $\mu_{\psi}$  produces
a sequence $(q_1, q_2, \ldots)$ that almost surely should have typical properties with respect to $\mu^{\infty}_{\psi}$.
In particular, such a sequence should typically be 1-random (in a suitable biased sense fixed by  $\mu_{\psi}$).\footnote{\label{noloss} We only defined 1-randomness for outcome sequences of fair coin flips, but  the definition can be extended to other measures, see Downey \& Hirschfeldt (2010), \S 6.12 and references therein.}   Anything remotely deterministic, like computable samplings, will only contribute  sequences that are atypical (i.e.\ collectively have measure zero) for $\mu_{\psi}$.\footnote{\label{31}
In this light I  draw attention to an important result of Senno (2017), Theorem 3.2.7, see also  Bendersky \emph{et al} (2017), which is entirely consistent with the above analysis: If  the functions $f_1$ and $g$  are computable (within a computable time bound), then Alice and Bob can signal superluminally.
In other words, where mathematically speaking averaging the hidden state over the probability measure  $\mu_{\psi}$ suffices to guarantee Surface Locality even in a theory without Hidden Locality (Valantini, 2002), if this averaging is done by sampling $Q$ in a long run of repeated measurements, then this sampling must at least be incomputable.} 

Likewise for the sampling of $X$ with respect to $\mu_{\psi}'$ in 't Hooft's theory.
Thus the requirement that the functions $f_1$ and $g$ randomly sample $\mu_{\psi}$ and $\mu_{\psi}'$ 
 introduces an element of unavoidable randomness into the hidden variable theories in question, which seems whackingly at odds with their deterministic character. 
 Indeed, I only see two possibilities:
\begin{itemize}
\item  This sampling is provided by the hidden variable theory. In that case, the above argument shows that the theory must contain an irreducibly random ingredient.
\item  The sampling is not provided by the theory. In that case, the theory fails to determine the outcome of any specific experiment and just provides averages of outcomes.
\end{itemize}
Either way, although at first sight our hidden variable theories are (Laplacian) deterministic (as is quantum mechanics, see footnote \ref{STE}), in their accounting for measurement outcomes they are not (again, like quantum theory). 
What is the source of indeterminism?
\begin{itemize}
\item  In standard (Copenhagen) \qm\  this source lies in the outcomes of experiments given the quantum state, whose associated Born measure is sampled;
\item In ``deterministic" hidden variable theories it is the assignment of the hidden variable to each measurement no.\ $n$, i.e.\ the sampling of the Born-like measures $\mu_{\psi}$ and $\mu_{\psi}'$.
\end{itemize}
So at best the source  of indeterminism has been shifted. Moreover, in Bohmian mechanics and 't Hooft's theory $\mu_{\psi}$ and $\mu_{\psi}'$ 
 \emph{equal} the Born measure, so one wonders what has been gained against Copenhagen \qm.
Therefore, one has to conclude \vspace{5pt} that:

\noindent \emph{\textbf{Truly} deterministic hidden variable theories (i.e.\ those in which well-defined experiments have determined outcomes \emph{given} the initial state \textbf{and} no appeal has to made to irreducibly random samplings  \emph{from} this state) compatible with the Born rule  do not exist}.\vspace{5pt}

\noindent
In other words,  as long as they reproduce all statistical predictions of \qm, deterministic theories underneath \qm\ still need a source of irreducible randomness in each and every experiment.  In my view, this defeats their purpose.\footnote{A related point about the double role of $\psi$ in  the pilot wave theory of De Broglie was made by Pauli  (1953, p.\ 38): `The hypothesis of a general probability distribution for the hidden variables that is determined by the single [wave] function $\psi$ is not justified from the point of view of a deterministic scheme: it is borrowed from a theory which is based on the totally different hypothesis that the [wave] function provides a complete description of the system.'
I am indebted to Guido Bacciagaluppi for this information.}

In classical coin tossing the role of the hidden state is played by  the initial conditions (cf.\ Diaconis \& Skyrms, 2018, Chapter 1, Appendix 2). The 50-50 chances (allegedly) making the coin fair are obtained by averaging over the initial conditions, i.e., by sampling. By the above argument, this sampling cannot be  deterministic, for otherwise the outcome sequences appropriate to a fair coin do not obtain: it must be done in a genuinely random way. This is  impossible classically, so that \emph{fair classical coins do not exist}, as  confirmed by the experiments of Diaconis \emph{et al} reviewed in Diaconis \& Skyrms (2018), Chapter 1.

In response to this argument, both the Bohmians and 't Hooft go for the second option and blame the randomness in question on the initial conditions,  whose specification is indeed  usually seen as lying outside the range of a deterministic theory.\footnote{The Bohmians are  divided on the origin of the quantum equilibrium distribution, cf.\ 
D\"{u}rr, Goldstein, \& Zanghi (1992),  Callender (2007), Norsen (2018), and  Valentini (2019).  The origin of $\mu_{\psi}$ or  is not my concern here; the problem I address is the need to randomly sample it and the justification for doing so.
} As explained by both parties 
(D\"{u}rr, Goldstein, \& Zanghi, 1992; 't Hooft, 2016), the randomness in the outcomes of measurement on quantum system, including the Born rule, is a direct consequence of the above randomness in initial conditions. But  in a Laplacian deterministic theory  one can either predict or retrodict and these procedures should be equivalent; so \emph{within the context of a deterministic hidden variable theory of the kinds under discussion}, Copenhagenists attributing the origin of randomness to the \emph{outcomes} of measurement and our hidden variable theorists attributing it to the \emph{initial conditions} for measurement, should be equivalent. Once again, this makes it impossible to regard the hidden variable theories in question as \emph{deterministic} underpinnings of (Copenhagen) \qm.

 Bohmians (but not 't Hooft!)  have another problem, namely 
  the function \er{Bsettings} that provides the  settings.  Although $f_2$ is outside their theory,  Bohmians 
  should either account for both the ``freedom" of choosing these settings and their randomness, or stop citing Bell's Theorem  (whose proof relies on averaging over random settings) in their favour.
  
 \noindent  Bell (1985) tried to kill these two birds with the same stone by saying that the settings had to be `at least effectively free for the purpose at hand', and clarifiying this as follows: 
\begin{quote}\begin{small}
Suppose that the instruments are set at the whim, not of experimental physicists, but of  mechanical random generators. (\ldots)  
Could the input of such mechanical devices be reasonably be regarded as sufficiently free for the purpose at hand? I think so. Consider the extreme case of a ``random" generator which is in fact perfectly deterministic in nature and, for simplicity, perfectly isolated. In such a device the complete final state perfectly determines the complete initial state--nothing is forgotten. And yet for many purposes, such a device is  precisely a ``forgetting machine". (\ldots) To illustrate the point, suppose that the choice between two possible [settings], corresponding to $a$ and $a'$, depended on the oddness of evenness of the digit in the millionth decimal place of some input variable. Then fixing $a$ or $a'$ indeed fixes something about the input--i.e., whether the millionth digit is odd or even. But this peculiar piece of information is unlikely to be the vital piece for any distinctly different purpose, i.e., it is otherwise rather useless. (\ldots) In this sense the output of such a device is indeed a sufficiently free variable for the purpose at hand. (Bell, 1985, p.\ 105).
\end{small}\end{quote}
This seems to mix up the two issues. Though independence of the settings is defensible in a theory like Bohmian mechanics (Esfeld, 2015), concerning their randomness 
Bell apparently ignored von Neumann's warning against mechanical (i.e.\ pseudo) random generators:
\begin{quote}\begin{small}
Any one who considers arithmetical methods of producing random digits is, of course, in a state of sin (von Neumann, 1951).
\end{small}\end{quote} 
See also Markowsky (2014). Thus Bell's statement was questionable already at the time of writing, but today we  know for sure that mechanical random generators leave a  loophole in the EPR--Bohm--Bell experiment: as soon as just one of the functions defining the settings (i.e.\ either of Alice or of Bob) is computable,\footnote{The computable function is subject to a time bound, but this is true for all pseudo-random generators.}
there is a model that is local (in the sense of Bell).
 See  Bendersky \emph{et al} (2016) or Senno (2017), Theorem 2.2.1
This  implies that Bohmian mechanics (as well as  other deterministic hidden variable theories that leaves the settings free) requires even more randomness than the sampling of the (Born) probability measure $\mu_{\psi}$, which further undermines the claim that it is a  deterministic theory.\footnote{\label{fnqrng} In order not to be in a state of sin even Copenhagen \qm\ has yet to prove that it is capable of building True Random Number Generators. Arguments in this direction are given in Abbott \emph{et al} (2012, 2019),  
 Ac\'{\i}n  \emph{et al} (2016), Bendersky  \emph{et al} (2016), Herrero-Collantes \& Garcia-Escartin (2017), and Kovalsky \emph{et al} (2018). See also the end of Appendix \ref{AR} for the (lacking) connection with 1-randomness.}
 
  The analysis given so far focused on the necessity of correctly sampling a probability measure $\mu$: if, so far in the context of hidden variable theories, where $\mu=\mu_{\psi}$, this is not done correctly, quantum-mechanical predictions such as Surface Locality may be threatened. But in general there \emph{are} measurement outcome sequences that fail to sample $\mu$:  atypical events with very low  or even zero probability can and do occur. This is even the whole point of the ``law of truly large numbers" quoted in \S\ref{RFR}!
 In general, the Hidden Locality (or no-signaling) property of \qm\ states that the probability
\begin{equation}
P_{\ps}(\al\mid a,b):= \sum_\beta P_{\ps}(\al, \beta\mid a, b) \label{NS}
\end{equation}
is independent of $b$, where $P_{\ps}(\al, \beta\mid a, b)$
is the \emph{Born probability} that measurement of observables determined by the settings $a$ and $b$ give outcomes $\al$ and $\beta$. Indeed, we have
\begin{equation}
P_{\ps}(\al\mid a,b)=P_{\ps_{|A}}(\al\mid a),
\end{equation}
where $\psi_{|A}$ is the restriction of the state $\psi$ on $B(H_A\ot H_B)$ to Alice's part $B(H_A)$. Similarly, the Born probability
$P_{\ps}(\beta\mid a,b)$ should be independent of $a$ and in fact equals $P_{\ps_{|B}}(\beta\mid b)$.

\noindent However, I have repeatedly noted that the \emph{empirical} probability extracted from a long measurement outcome sequence coincides with the corresponding \emph{Born} probability only almost surely with respect to 
the induced probability measure on the space of outcome sequences, and hence outliers violating the property \er{NS} exist (for finite sequences even with positive probability). If one such run is found, the door to superluminal signaling is  open, at least in principle.
To see this, recall that the crudest form of determinism is what is called \emph{Predictability} by Cavalcanti \& Wiseman (2012), i.e.\ the property that 
\beq
P_{\ps}(\al, \beta\mid a, b)\in\{0,1\}. \L{CW}
\eeq
 It is easy to show that the conjunction of Predictability and Surface Locality implies factorization and hence, for random settings, the Bell (CHSH) inequalities, and therefore for suitable states $\psi$ this conjunction contradicts the statistical predictions of \qm\ as expressed by the Born rule. Accepting the latter,  Surface Locality  therefore implies unpredictability and hence some (very weak) form of randomness. There are many other results in this direction,  ranging from e.g.\ Barrett, Kent, \& Hardy (2005) to Wolf (2015),\footnote{Note that the aim of  Barrett, Kent, \& Hardy (2005) is to prove security of some quantum key distribution protocol on the basis of Surface Locality \emph{even if \qm\ turns out to be incorrect}, whereas  the present paper investigates the role of statistical outliers \emph{assuming \qm\ is correct}. These aims are closely related, of course, since too many (how many?) outliers may make one question the theory.} involving varying definitions of randomness,
  all coming down to the implication 
 \begin{equation}
\mathit{no\: signaling} \Raw \mathit{randomness}, \label{former}
 \end{equation}
 assuming Freedom and random settings.\footnote{See Hermens (2019) for arguments for \er{former} and \er{latter} even without assuming Freedom.}
 What I'd like to argue for is the contrapositive 
  \begin{equation}
\mathit{no\: randomness} \Raw  \mathit{signaling}. \label{latter}
 \end{equation}
My argument is only heuristic, since the terms are used differently in  \er{former} and \er{latter}: to prove \er{former} one typically uses  Born probabilities and other theoretical entities, whereas for \er{latter} I use probabilities obtained from  outcome sequences as limiting relative frequencies: \er{latter} then comes from low-probability sequences that \emph{violate} \er{NS},
 which is \emph{satisfied} with 
probability one by sufficiently random sequences. Indeed, this difference is the whole point:
\begin{center}
 \emph{Surface Locality is a statistical property, like the second law of thermodynamics}.\footnote{
Similarly, Valentini (2002ab)  pointed out that violation of  the equilibrium condition in Bohmian mechanics leads to superluminal signaling. Giving an actual signaling protocol for an outcome sequence  (in the EPR--Bohm--Bell setting) that violates \er{NS} is highly nontrivial. 
 Yurtsever (2001) gives a protocol for superluminal signaling if the outcome sequence is not 1-random, but his arguments are tricky and the paper with complete details he announces has never appeared.  His approach may  be neither viable nor necessary, since such a protocol need only exist for outcome sequences violating \er{NS}, which is a \emph{stronger} property than violating 1-randomness--since contrapositively 1-randomness implies all averaging properties like \er{NS}. In fact, further assumptions are necessary in order for Alice to have a chance (\emph{sic}) to learn Bob's settings $b$ from her part $(\al_n)_n$ of the (double) outcome sequence $(\al_n,\beta_n)_n$; a complete lack of structure would also make any kind of signaling random. I know of only one such protocol (which, however,  should  suffice as a ``proof of concept"): if the unlikely outcome sequence $(\al_n,\beta_n)_n$ comes from \emph{computable} functions $\al_n=\al(a,b,n)$ and $\beta_n=\beta(a,b,n)$ with $O(T^2)$ time bounds, then signaling is possible through a learning algorithm for computable functions, see Senno (2017), Theorem 3.2.7 and Bendersky \emph{et al} (2017). This is the same result as in footnote \ref{31}; by Proposition 3.1.7 in Senno (2017), \emph{mutatis mutandis} it applies both with and without hidden variables.
 This protocol enables Alice (or Bob) to signal after a \emph{finite} (but alas unknown and arbitrarily large) number of of measurements (this answers a speculation by
 Carlo Rovelli that  finite-size corrections to the lack of randomness may conspire to prevent superluminal signaling).
 }
\end{center}
 \appendix
 \section{The Born rule revisited}\label{BR}
 Every argument on probabilities in \qm\ relies on the Born rule, which is so natural that it will probably (!) be with us forever. In order to explain its canonical mathematical origin, I use the C*-algebraic approach to quantum theory, which because of its ability to simultaneously talk about commutative and non-commutative algebras (and hence about classical and quantum probabilities) is especially useful in this context.\footnote{See e.g.\ Haag (1992), Ruetsche (2011), or Landsman (2017) for the  C*-algebraic approach. My discussion of the Born rule is partly taken from the latter, which in turn elaborates on Landsman (2009), but the corresponding ``classical" construction has been added and may be new; I find it very instructive.} Indeed, the Born rule arises  by restricting a quantum state $\om$ on the non-commutative algebra $B(H)$, i.e.\ the algebra of all bounded operators on some \Hs\ $H$,  to the commutative  C*-algebra $C^*(a)$ generated by some self-adjoint operator $a=a^*\in B(H)$ and the unit operator $1_H$ (so perhaps $C^*(a,1_H)$ would have been better notation).\footnote{This is well defined, since $C^*(a)$ is  the intersection of all C*-algebras in $B(H)$ that contain $a$ and $1_H$.}  
 
 This view has been inspired by  the Copenhagen Interpretation, in that the associated probabilities originate in the classical description of a measurement setting, as called for by Bohr. However, 
precisely because of this, my derivation also seems to undermine the Copenhagen claim of the irreducibility of randomness, which may now clearly be traced back to a voluntary loss of information in passing from an algebra to one of its subalgebras and ignoring the rest (see also \S\ref{RQM}). Happily, the Copenhagen Interpretation will be saved from this apparent inconsistency by a mathematical property that has no classical analog.

As I see it, the Born rule is the quantum-mechanical counterpart of the following elementary construction in measure theory. 
Let $X$ be a compact Hausdorff space,\footnote{With some modifications the construction easily generalizes to the locally compact case.} with associated commutative \ca\ $A=C(X)$ of complex-valued continuous functions on $X$. 
Let $f\in A$ be real-valued and denote its (automatically closed) range in $\R$ by $\sg(f)$; the notation is justified by the fact that the spectrum of a self-adjoint operator in quantum theory is analogous to $\sg(f)$ in every possible way. Let $\om$ be a state on $A$, or, equivalently (by the Riesz representation theorem), a probability measure $\mu$ on $X$, related to $\om$ by 
\beq
\om(f)=\int_X d\mu(x)\, f(x).
\eeq 
The state $\om$ then also defines a probability measure $\mu_f$ on the ``spectrum" $\sg(f)$ through
  \begin{equation}
\mu_f(\Dl)=\mu(f\in \Dl), \label{muf}
\end{equation}
where  $\Dl\subseteq \sg(f)$ and $f\in\Dl$ or $f\inv(\Dl)$ denotes the subset $\{x\in X\mid f(x)\in\Dl\}$ of $X$. Virtually all of probability theory is based on this construction, which I now rephrase. For any C*-algebra $A$ with unit $1_A$, let $C^*(a)$ be the smallest C*-algebra in $A$ (under the same operations and norm) that contains $a$ and the unit $1_A$, as above for $A=B(H)$.
 If $a^*=a$, then $C^*(a)$ is the norm-closure of the set of all (finite) polynomials in $a$.
Take $A=C(X)$  and $f\in C(X)$, assumed real-vaued (since $f^*=f$). This gives as isomorphism
\begin{align}
C(\sg(f))&\stackrel{\cong}{\raw} C^*(f); \label{Cstarf1}\\
g&\mapsto g\circ f,\label{Cstarf2}
\end{align} 
where $f\in C(X)$, $g\in C(\sg(f))$, as follows from  the Stone--Weierstrass Theorem.\footnote{
if $K\subset\R$ is compact, then polynomials on $K$ are dense in $C(K)$ for the supremum-norm, i.e.\ the norm on $C(X)$. To prove  that \er{Cstarf1} - \er{Cstarf2} is an isomorphism of C*-algebras one may therefore start with polynomial functions $g$ (taking $K=\sg(f)$) and finish with a continuity argument.}

  If $C$ is a C*-subalgebra of $A$ with the same unit as $A$, then any state $\om$ on $A$ defines a state $\om_{|C}$ on $C$. Thus $\om_{|C^*(f)}$ is a state on $C^*(f)$. Furthermore, if $\phv:B\raw C$ is a unital homomorphism of unital C*-algebras (i.e., a linear map preserving all operations as well as the unit---in our case $\phv$ is even an isomorphism), and if $\om'$ is a state on $C$, then $\phv_*\om'= \om'\circ \phv$ is a state on $B$. Using  \er{Cstarf1} - \er{Cstarf2}, we apply this construction to
\begin{align}
A=C(X), && B=C(\sg(f)), && C= C^*(f); \label{CBM1} 
\\  \om\in S(C(X)), && \om'=\om_{|C^*(f)}, && \phv(g)=g\circ f,\label{CBM2}
\end{align}
where $S(A)$ is the state space of $A$,  so that a state $\om$  on $C(X)$  defines a state $\om'$ on $C(\sg(f))$. Once again using the Riesz representation theorem then turns $\om'$ into a probability measure $\mu'$ on $\sg(f)$;  \emph{and this is exactly the probability measure 
 $\mu_f=\mu'$ defined in \er{muf}}.
 
Copying this reasoning for \qm\  \emph{mutatis mutandis} immediately gives the Born measure. Instead of \er{CBM1} - \er{CBM2}, for some given $a=a^*\in B(H)$ we now take
\begin{align}
A=B(H), && B=C(\sg(a)), && C= C^*(a); \label{QBM1} 
\\  \om\in S(B(H)), && \om'=\om_{|C^*(a)}, && \phv(g)=g (a),\label{QBM2}
\end{align}
where  \er{Cstarf1} - \er{Cstarf2} is replaced by the  continuous functional calculus (CFC), that is, 
\begin{align}
C(\sg(a))&\stackrel{\cong}{\raw} C^*(a); \label{CFC1}\\
g&\mapsto g(a).\label{CFC2}
\end{align} 
Here $\sg(a)$ is the spectrum of $a$, defined as the set of $\lm\in\C$ for which the operator $a-\lm\cdot 1_H$ is \emph{not} invertible in $B(H)$; if $\dim(H)<\infty$, then $\sg(a)$ is the set of eigenvalues of $a$. In most applications the state $\om$ is \emph{normal}, i.e., given by a density operator $\rh$ on $H$ through
\begin{equation}
\om(a)=\Tr(\rh a).  \label{NState}
\end{equation}
We then replay our record: a state $\om\in S(B(H))$ restricts to a state $\om_{|C^*(a)}$ on $C^*(a)$, which is mapped to a state on $C(\sg(a))$ by the CFC, which state in turn is equivalent to a  probability measure on $\sg(a)$.
The probability measure $\mu_a$ on $\sg(a)$ obtained by this construction is exactly the Born measure, which is  more commonly defined as follows:
 \begin{theorem}\label{defBornmu}
 Let $H$ be a \Hs, let $a^*=a\in B(H)$, and let $\om$ be a state on $B(H)$.
  There exists a unique probability measure 
 $\mu_a$ on the spectrum $\sg(a)$ of $a$ such that 
 \begin{equation}
\om(g(a))=\int_{\sg(a)} d\mu_a(\lm)\, g(\lm), \:\: g\in C(\sg(a)).\label{BornfromGelfand}
\end{equation}
  \end{theorem}
We now return to the issue raised in the introduction to this appendix: even short of hidden variables (see \S\ref{RQM}), to what extent is the probabilistic structure of \qm\ that is already inherent in the Born measure  reducible to ignorance? From a naive perspective it is, for in the corresponding classical case the measure $\mu_f$ on $\sg(f)$ does not determine the measure $\mu$ on $X$ except when $C^*(f)=C(X)$ (which is certainly possible, take e.g. $X=[0,1]$ and $f(x)=x$). However, in the quantum case it so happens that if $a$ is \emph{maximal} (i.e.\ its spectrum is nondegenerate, or, equivalently, $C^*(a)$ is a maximal commutative C*-subalgebra of $B(H)$), then the measure $\mu_a$ on $\sg(a)$ \emph{does} determine the state $\om$, at least if $\om$ is pure and normal, as in \er{NState}, \emph{despite the fact that $C^*(a)$ is just a small part of $B(H)$}.
So the Copenhagen Interpretation is walking a tightrope here, but it doesn't fall.\footnote{This discussion is closely related to the so-called Kadison--Singer conjecture in operator algebras, which however is nontrivial only for non-normal states. 
See Stevens (2016) and Landsman (2017), \S 2.6 and \S 4.3  }

   The Born measure is a mathematical construction; what is its relationship to experiment? This relationship must be the source of the (alleged) randomness of \qm, for the Schr\"{o}dinger equation is deterministic. We 
start by postulating, as usual, that $\mu_a(A)$ is the (single case) probability that measurement of the observable $a$ in the state $\om$ (which jointly give rise to the pertinent Born measure $\mu_a$) gives a result $\lm\in A\subset\sg(a)$.  What needs to be clarified in the above statement is the word \emph{probability}. 

Although I do not buy Lewis's (1994) ``Best Systems Account" (BSA) of laws of nature,\footnote{See  Loewer (2004) and, concisely, Callender (2007) for summaries of Lewis's BSA.  As I understand it, Lewisian laws \emph{describe} nature, but fail to \emph{govern} or \emph{guide} it in the way that the Law of Moses, carved in stone, is supposed to influence (decent) human behaviour. Now I agree with  Loewer (2004) that it is hard to see how laws as mathematical expressions in physics books 
could govern anything, but I expect this problem to be solved in due course by  an 
 emergentist account in which lower-level behaviour--though never fundamental--gives rise to emergent laws at some higher substratum of reality. 
 See also footnote \ref{Emfn}.
} I do agree with his identification of single-case probabilities as numbers (consistent with the probability calculus as a whole) that \emph{theory} assigns to events, upon which long-run frequencies provide \emph{empirical evidence} for the theory in question,\footnote{Within the Lewisian ideology this is ultimately justified by his \emph{Principal Principle}, 
which roughly speaking  equates chance as subjective degree of belief (i.e.\ credence) with objective chance.}
 but do not \emph{define} probabilities.\footnote{This, as is well known by now, is a dead end street (H\'{a}jek \&  Hitchcock, 2016). It should be mentioned though that the fascinating yet ultimately flawed attempts of von Mises to provide such a definition played a decisive role in the road towards algorithmic randomness (van Lambalgen, 1987).} 
The Born measure is a case in point: these probabilities are \emph{theoretically given},  but have to be \emph{empirically verified} by long runs of
 independent experiments. In other words, by the results reviewed below such experiments provide numbers whose role it is to test the Born rule as a hypothesis. This is justified by \er{nuas5} in \S\ref{PP}, which equates probabilities \emph{computed} from long-run frequencies with the corresponding single-case probabilities that \emph{define} the clause ``with probability one" under which \er{nuas5} holds, and without which clause the limit in  \er{nuas5} would be undefined. 
As explained in \S\S\ref{PP}-\ref{CAC}, the relevance of   \er{nuas5} to quantum physics comes from Theorem \ref{ET}, which I now prove. 
\smallskip

  \emph{Proof of Theorem \ref{ET}}.
Let $a=a^*\in B(H)$, where $H$ is a \Hs\  and $B(H)$ is the algebra of all bounded operators on $H$, and let $\sg(a)$ be the spectrum of $a$. For simplicity (and since this is enough for our applications, where $H=\C^2$) I assume $\dim(H)<\infty$, so that $\sg(a)$ simply consists of the eigenvalues $\lm_i$ of $a$ (which may be degenerate). Let us first consider a \emph{finite} number $N$ of runs of an identical measurement of $a$.
The first option in the theorem corresponds to a simultaneous measurement of the commuting operators 
\begin{align}
a_1&=a\ot 1_H\ot\cdots\ot 1_H; \label{a1} \\
& \cdots \nn \\
a_N &=1_H\ot\cdots\ot 1_H\ot a, \label{aN}
\end{align}
all defined on the $N$-fold tensor product $H^N\equiv H^{\ot N}$ of $H$ with itself.\footnote{This can even be replaced by a single measurement, see Landsman (2017), Corollary A.20.}  To put this in a broader perspective,  consider \emph{any} set $(a_1, \ldots, a_N)\equiv\ul{a}$ of commuting operators on \emph{any} \Hs\ $K$ (of which \er{a1} - \er{aN} is obviously a special case with $K=H^N$).
These operators have a \emph{joint spectrum} $\sg(\ul{a})$, whose elements are the \emph{joint eigenvalues} $\ul{\lm}=(\lm_1, \ldots, \lm_N)$, defined by the property that there exists a nonzero joint eigenvector $\ps\in K$ such that $a_i\ps=\lm_i\ps$ for all $i=1, \ldots, N$; clearly, the joint spectrum is given by
\begin{equation}
\sg(\ul{a})=\{\ul{\lm}\in
\sg(a_1)\x\cdots\x\sg(a_N)\mid e_{\ul{\lm}}\equiv e^{(1)}_{\lm_1}\cdots e^{(n)}_{\lm_n} \neq 0\} \subseteq \sg(a_1)\x\cdots\x\sg(a_N), \label{21}
\end{equation}
where $e^{(i)}_{\lm_i}$ is the spectral projection of $a_i$ on the eigenspace for the eigenvalue $\lm_i\in\sg(a_i)$.
Von Neumann's Born rule for the probability $p_{\ul{a}}(\ul{\lm})$ of finding $\ul{\lm}\in\sg(\ul{a})$ then simply reads 
\begin{equation}
p_{\ul{a}}(\ul{\lm})=\om(e_{\ul{\lm}}),\label{paQn}
\end{equation}
 where $\om$ is the state on $B(K)$ with respect to which the Born probability is defined. If $\dim(K)<\infty$, as I assume, we always have $\om(a)=\Tr(\rh a)$ for some density operator $\rh$, and for a  general \Hs\ $K$ this is the case iff the state $\om$ is normal on $B(K)$. For (normal) pure states we have $\rh=|\psi\ra\la\psi|$ for some unit vector $\ps\in K$, in which case
 \begin{equation}
p_{\ul{a}}(\ul{\lm})=\la\ps,e_{\ul{\lm}}\psi\ra.\label{paQnp}
\end{equation}
The  Born rule \er{paQn} follows from the same reasoning as the single-operator case explained in the run-up to Theorem \ref{defBornmu} (Landsman, 2017, \S 4.1):\footnote{Moreover, quite apart from te above derivation, on separable \Hs s the  Born rule for  multiple commuting operators is in fact a special case of the one for single operators  (Landsman, 2017, \S 2.5).}
we have an isomorphism
\begin{equation}
C^*(\ul{a})\cong C(\sg(\ul{a}))
\end{equation}
of (commutative) C*-algebras, and under this isomorphism the restriction of the state $\om$, originally defined on $B(K)$, to $C^*(\ul{a})$ defines a probability measure $\mu_{\ul{a}}$ on the joint spectrum $\sg(\ul{a})$, which is just the Born measure whose probabilities are given by \er{paQn}.
For  \er{a1} - \er{aN} we have equality in \er{21}; since in that case $\sg(a_i)=\sg(a)$,  we obtain
\begin{equation}
\sg(\ul{a})=\sg(a)^N,
\end{equation}
and, for all $\lm_i\in\sg(a)$ and states $\om$ on $B(H^N)$, 
 the Born rule \er{paQn} becomes 
 \begin{equation}
p_{\ul{a}}(\lm_1, \ldots, \lm_N)=\om(e_{\lm_1}\ot\cdots\ot e_{\lm_N}).
\end{equation}
Now take a given state $\om_1$ on $B(H)$. Reflecting the idea that $\om$ is the state on $B(H^N)$ in which $N$ independent measurements of $a\in B(H)$ in the state $\om_1$ are carried out,
choose
 \beq
 \om=\om_1^N,
 \eeq
 i.e.\  the state on $B(H^N)$ defined by linear extension of its action on elementary tensors:
\beq
\om_1^N(b_1\ot\cdots\ot b_n)=\om_1(b_1) \cdots \om_N(b_N),
\eeq
where $b_i\in B(H)$ for each $i=1, \ldots, N$.
It follows that
\begin{equation}
\om^N(e_{\lm_1}\ot\cdots\ot e_{\lm_N})=\om_1(e_{\lm_1})\cdots\om_1(e_{\lm_N})=p_a(\lm_1)\cdots p_a(\lm_N),
\end{equation}
so that the joint probability of the outcome $(\lm_1, \ldots, \lm_N)\in\sg(\ul{a})$ is simply
\begin{equation}
p_{\vec{a}}(\lm_1, \ldots, \lm_N)=p_a(\lm_1)\cdots p_a(\lm_N).
\end{equation}
Since these are precisely the probabilities for option 2 (i.e.\ the Bernoulli process), i.e.,
 \beq
 \mu_{\ul{a}}=\mu_a^N, \label{mulaN}
 \eeq
 this
 proves the claim for $N<\infty$.  
 
 To describe the limit $N\raw\infty$, let $B$ be any C*-algebra with unit $1_B$;  below I take $B=B(H)$, $B=C^*(a)$,  or $B=C(\sg(a))$. 
 We now take
 \beq
 A_N=B^{\ot N},
 \eeq
  the $N$-fold tensor product 
  of $B$ with itself.\footnote{If $B$ is infinite-dimensional, for technical reasons  the so-called \emph{projective} tensor product should be used.} The  special cases above may  be rewritten as
\begin{align}
B(H)^{\ot N}&\cong  B(H\ot \cdots \ot H);\label{BHN}\\
C^*(a)^{\ot N}&\cong C^*(a_1, \ldots, a_N);\label{StaN} \\
C(\sg(a))^{\ot N}&\cong  C(\sg(a) \x\cdots\x\sg(a)), \label{sgN}
\end{align}
 with $N$ copies of $H$ and $\sg(a)$, respectively; in \er{StaN} the $a_i$ are given by \er{a1} - \er{aN}.
 
 We may then wonder if these algebras have a limit as $N\raw\infty$. They do, but it is not unique and depends on the choice of observables, that is, of the infinite sequences $\mathbf{a}=(\mathsf{a}_1,\mathsf{a}_2, \ldots)$, with $\mathsf{a}_N\in A_N$, that are supposed to have a limit. One possibility is to take sequences $\mathbf{a}$ for which  there exists $M\in\N$ and $\mathsf{a}_M\in A_M$ such that for each $N\geq M$,
\beq
\mathsf{a}_N=\mathsf{a}_M \ot 1_B\cdots\ot 1_B, \label{anam}
\eeq
 with $N-M$ copies of the unit $1_B$. On that choice, one obtains the infinite tensor product $B^{\ot\infty}$, see Landsman (2017), \S C.14. The limit of \er{BHN} in this sense is $B(H^{\ot\infty})$, where $H^{\ot\infty}$ is von Neumann's `complete' infinite tensor product of \Hs s,\footnote{See Landsman (2017), \S 8.4 for this approach. The details are unnecessary here.}  
in which $C^*(a)^{\ot\infty}$ is the C*-algebra generated by the operators $(a_1,a_2, \ldots)$. 
 The limit of \er{sgN} is 
 \begin{equation}
C(\sg(a))^{\ot \infty}\cong C(\sg(a)^{\N}), 
\end{equation}
where $\sg(a)^{\N}$, which we previously  saw as a measure space (as a special case of $X^{\N}$ for general compact Hausdorff spaces $X$), is now seen as a topological space with the product  topology, in which it is compact.\footnote{From Tychonoff's Theorem. The associated Borel structure is the one defined in footnote \ref{sigma} in \S\ref{CAC}.} As in the finite case, we  have an isomorphism
\begin{equation}
C^*(a)^{\ot\infty}
\cong C(\sg(a))^{\ot \infty},
\end{equation}
and hence, on the given identifications and the notation \er{a1} - \er{aN}, we have
\begin{equation}
C^*(a_1, a_2, \ldots)\cong C(\sg(a)^{\N}). \label{a1a2iso}
\end{equation}
It follows from the definition of the infinite tensor products used here that each state $\om_1$ on $B$ defines a state
$\om_1^{\infty}$ on $B^{\ot\infty}$. Take $B=B(H)$ and restrict $\om_1^{\infty}$, which \emph{a priori} is a state on $B(H^{\ot\infty})$, to its commutative C*-subalgebra $C^*(a_1, a_2, \ldots)$. The isomorphism \er{a1a2iso} then gives a probability measure $\mu_{\ul{a}}$ on the compact space $\sg(a)^{\N}$, where the label $\ul{a}$ now refers to the infinite set of commuting operators $(a_1, a_2, \ldots)$ on $H^{\ot\infty}$. To compute this measure, use  \er{BornfromGelfand} and the fact that by construction functions of the type
\beq
f(\lm_1, \lm_2, \ldots)=f^{(N)}(\lm_1, \ldots, \lm_N),
\eeq  
 where $N<\infty$ and $f^{(N)}\in C(\sg(a)^N)$, are dense in $C(\sg(a)^{\N})$ with respect to the appropriate supremum-norm, and that in turn finite linear combinations of factorized functions $f^{(N)}(\lm_1, \ldots, \lm_N)=f_1(\lm_1)\cdots f_N(\lm_N)$ are dense in $C(\sg(a)^{N})$. 
 It follows from this that 
 \beq
 \mu_{\ul{a}}=\mu_a^{\infty}. \label{mula}
 \eeq
 Since this generalizes \er{mulaN} to $N=\infty$, 
 this finishes the proof of Theorem \ref{ET}. \hfill $\Box$
 \section{1-Randomness}\label{AR}
 As we have seen, randomness (of measurement outcomes) in \qm\ was originally defined by Born as \emph{indeterminism}. This is  what hidden variable theories like 't Hooft's and Bohmian mechanics challenge, to which in turn Bell's Theorem and the Free Will Theorem (read in a specific way reviewed in Appendix \ref{FWT} below) provide obstacles. Indeterminism is a physical definition of randomness, which in mathematics is most closely matched by \emph{incomputability}. However, there is a much deeper notion of randomness in mathematics, which is what at least in my view \qm\  should aspire to produce. 
This notion, now called \emph{1-randomness}, has its roots in Hilbert's Sixth Problem, viz.\ the \emph{Mathematical Treatment of the Axioms of Physics},\footnote{`The investigations on the foundations of geometry suggest the problem: To treat in the same manner, by means of axioms, those physical sciences in which mathematics plays an important part; in the first rank are the theory of probabilities and mechanics.' (Hilbert, 1902, p.\ 448). See also Gorban (2018).} which was taken up independently and very differently  by von Mises (1919) and by Kolmogorov (1933). 

 Von Mises was a strict frequentist for whom probability was a derived concept, predicated on first having a good notion of a random sequence from which relative frequencies \emph{defining} probability could be extracted. Kolmogorov, on the other hand, started from an axiomatic \emph{a priori} notion of probability from which a suitable mathematical concept of randomness was subsequently to be extracted. Despite the resounding and continuing success of the first step, 
Kolmogorov's initial failure to achieve the follow-up  led to his later notion of algorithmic randomness, which (subject to a technical improvement) was to become one of the three equivalent definitions of 1-randomness. In turn, von Mises's failure to adequately define random sequences eventually led to the other two definitions.\footnote{See van Lambalgen (1987) or, for a lighter account, Diaconis \& Skyrms (2018), for this history. }  

Kolmogorov's problem, which was noticed already by Laplace and perhaps even earlier probabilists, was that, specializing to a 50-50 Bernoulli process for simplicity,\footnote{A \emph{string} $\sg$ is finite row of bits, whereas a \emph{sequence} $x$ is an infinite one.
 The length of a string $\sg$ is $|\sg|$.} 
  any  binary string $\sg$ of length $N$ has probability $P(\sg) = 2^{-N}$ and any (infinite) binary sequence $x$ has probability $P(x) = 0$, although say $\sg=0011010101110100$ looks much more random than $\sg=111111111111111$. In other words, their \emph{probabilities} say little or nothing about the \emph{randomness} of individual outcomes. Imposing statistical properties helps but is not enough to guarantee randomness. It is slightly easier to explain this in base 10, to which I therefore switch for a moment. If we call a sequence $x$  \emph{Borel normal} if each possible string $\sg$ in $x$ has (asymptotic) frequency $10^{-|\sg|}$ 
    (so that each digit $0, \ldots, 9$ occurs 10\% of the time,  each block $00$ to $99$ occurs 1\% of the time, etc., then   \emph{Champernowne's number}
    $$ 0123456789101112131415161718192021222324252629282930 \ldots$$
 can be shown to be Borel normal. The decimal expansion of $\pi$ is conjectured to be Borel normal, too (and has been empirically verified to be so in billions of decimals), but these numbers are hardly random: they are computable,  which is an antipode to randomness. 
 
 Von Mises's problem was that his definition of a random sequence (called a \emph{Kollektiv}), despite his great insights, simply did not work. Back to binary sequences, his 
 notion of randomness was supposed to guarantee the existence of limiting relative frequencies (which in turn should define the corresponding single-case probabilities), but of course he understood that a sequence like $01010101\cdots$ is not very random at all. His idea was that  limiting relative frequencies should exist not only for the given sequence, but also for all its permutations defined by so-called \emph{place-selection functions}, which he tried to find (in vain) by precluding successful gambling strategies on the digits of the sequence.
 
\noindent  Of the three equivalent definitions of 1-randomness  already mentioned in \S\ref{RFR}, namely:
  \begin{center}
\emph{Incompressibility}; \hspace{1.5cm}
 \emph{Patternlessness};\hspace{1.5cm} 
 \emph{Unpredictability},
\end{center}
 the first may be said to go back to Kolmogorov's struggle, whereas the other two originate in later attempts by other mathematicians to improve the work of von Mises.\footnote{As mentioned and referenced in most literature as well as in the next footnote, Kolmogorov's work in the 1960s on the first definition was predated by Solomonoff and matched by independent later work of  Chaitin. Key players around the other definitions were Schnorr and  Martin-L\"{o}f, respectively. }

  Although there 
  isn't a single ``correct" mathematical notion of randomness, the notion of 1-randomness featured here stands out (and represents a consensus view) largely because 
 it can be defined in these three equivalent yet fairly different ways, each of which realizes some basic intuition on randomness (of course defined through its obvious antipode!). 
 
 In what follows,  these notions will be defined more precisely, followed by some of their consequences.\footnote{In increasing order of technicality, readers interested in more detail are referred to Diaconis \& Skyrms (2018, Chapter 8), Terwijn (2016), or Volchan (2002) at a popular level, then Gr\"{u}nwald \& Vit\'{a}nyi (2008), Dasgupta (2011),   Downey \emph{et al} (2006), or Muchnik \emph{et al} (1998), then Li \& Vit\'{a}nyi (2008) or Calude (2010), and finally Downey \& Hirschfeldt (2010).
See also  Baumeler  \emph{et al} (2017),  Bendersky \emph{et al}  (2014), 
Calude (2004),  Eagle  (2019), Earman (2004), Kamminga (2019), Senno (2017), Svozil (1993, 2018),  Wolf (2015), and Zurek (1989) for brief introductions to Kolmogorov randomness with  applications to  physics.} We assume basic familiarity with the notion of a computable function $f:\N\raw\N$, which may technically be defined  through recursion theory or equivalently through Turing machines: \emph{a function is computable if it can  be computed by a computer}.    
\subsection*{Incompressibility}
The idea is that a string or sequence is random iff its shortest description is the sequence itself, but the notion of a description has to made precise to avoid \emph{Berry's paradox}:
\begin{quote}\begin{small}
The Berry number is the smallest positive integer that cannot be described in less than eighteen words.
\end{small}\end{quote}
The paradox, then, is that on the one hand this number must exist, since only finitely many integers can be described in less than eighteen words and hence the set of such numbers must have a lower bound, while on the other hand Berry's number cannot exists by its own definition. This is, of course, one of innumerable paradoxes of natural language, which, like the liar's paradox, will be seen to lead to an incompleteness theorem once the notion of a description has been appropriately formalized in mathematics, as follows.\footnote{We write $\ul{2}^{\ul{N}}$ for the set of all binary strings $\sg$  of length $|\sg|=N\in\N$, and $\ul{2}^*=\cup_N\ul{2}^{\ul{N}}$ for the set of all binary strings. Finally, 
  $\ul{2}^{\N}$ denotes the set of all binary sequences $x$  (which are infinite by convention). }

The \emph{plain Kolmogorov complexity}  $C(\sg)$ of $\sg\in \ul{2}^{\ul{N}}$ is defined as the length (in bits) of the shortest computer program (run on some fixed universal Turing machine $U$) that computes $\sg$. The choice of $U$ affects this definition only up to a $\sg$-independent constant. For technical reasons (especially for defining the randomness of sequences) it is preferable to work with \emph{prefix-free Turing machines} $T$, whose domain $D(T)$ consists of a prefix-free subset of $\ul{2}^*$, i.e., if $\sg\in D(T)$ then $\sg\ta\notin D(T)$ for any $\sg,\ta\in \ul{2}^*$, where $\sg\ta$ is  the concatenation of $\sg$ and $\ta$, as the notation suggests. This is also independent of $U$ up to a constant, and defines the
the \emph{prefix-free Kolmogorov complexity} $K(\sg)$ as  the length of the shortest  program (run on some fixed universal prefix-free Turing machine $U$) that computes $\sg$. 
For fixed  $c\in\N$ we then say that $\sg\in \ul{2}^*$ is \emph{$c$-compressible} if $K(\sg)<|\sg|-c$; of course, this  depends on $U$ via $K(\sg)$, but nonetheless a simple counting argument shows:
\begin{center}
 At least $2^N-2^{N-c+1}+1$ strings $\sg$ of length $|\sg|=N$ are 
\emph{not} $c$-compressible. 
\end{center}
Clearly, as $\sg$ grows and $c>0$, these will form the overwhelming majority. Finally, $\sg$ is
\emph{Kolmogorov $c$-random} if it is not  $c$-compressible, i.e., if $K(\sg)\geq |\sg|-c$, and 
\emph{Kolmogorov random} if it is not  $c$-compressible for any $c\in\N$, that is, if one even has $K(\sg)\geq |\sg|$. Since $K(\sg)$ can at most be equal to the length of an efficient printing program plus the length of $\sg$, this simply means that $\sg$ is Kolmogorov random if 
\beq
K(\sg)\approx |\sg|,
\eeq
where  $\approx$ means `up to a $\sg$-independent constant'. Finally, if $x_{|N}$ is the initial segment of length $N$ of 
a sequence $x\in\ul{2}^{\N}$, i.e.\  $|x_{|N}|=N$, then $x$ is  \emph{Kolmogorov--Chaitin random} if
\begin{equation}
\lim_{N\raw\infty} \frac{K(x_{|N})}{N}=1.\label{Calude}
\end{equation}
Equivalently,\footnote{See Calude (2010), Theorem 6.38 (attributed to Chaitin) for this equivalence.}
 a sequence $x$ is Kolmogorov--Chaitin random if there exists $c\in\N$ such that each truncation $x_{|N}$ 
satisfies   $K(x_{|N})\geq N - c$.  I note with satisfaction that randomness of sequences, seen as idealizations of finite strings, as such satisfies Earman's principle (\S\ref{PP}).

This definition of randomness of both finite strings and infinite sequences looks very appealing, but in a way it is self-defeating, since although it is defined in terms of computability, the  complexity function $K$ is not computable, and hence one cannot even determine algorithmically if  a finite string $\sg$ is random (let alone an infinite one).\footnote{To see this,  define a function $L:\N\raw\N$ by 
$L(n) = \min\{m\in\N \mid K(m) \geq n\}$. Identifying $\N\cong \ul{2}^*$ (in a computable way),
$L(n)$, seen as an element of $\ul{2}^*$,  is  the shortest string $m$ whose complexity $K(m)$ exceeds $n$, so that  by construction $K(L(n)) \geq n$.  If $K$ were  computable, then so would $L$ be. Suppose $T$ is the shortest prefix-free program  that computes $L$. 
Since $K(L(n))$ is the length of shortest prefix-free program that computes $L(n)$, we  have
  $K(L(n)) \leq |n| + |T|$, where $|n|$ is the length of $\sg$ under the above bijection $n\lraw\sg$,
 so that $|n|\approx \mbox{}_{10}\!\log n$.
  Thus  $n\leq  K(L(n)) \leq |n| + |T|$, which cannot be true for large $n$.}
  
Moreover, Berry's paradox strikes back through \emph{Chaitin's incompleteness Theorem}: For any consistent, sound, and sufficiently rich formal system $F$ (containing basic arithmetic) there is a constant $f$ such that the statement $K(\sg) > f$ cannot be proved in $F$ for any $\sg\in \ul{2}^*$, although it is true for  all (random)  strings $\sg$ that satisfy $K(\sg) \geq|\sg| > f$ (and there are infinitely many such strings, as the above counting argument shows).
 To get a very rough idea of the proof, let me just say that 
any proof in $F$ of the sentence $K(\sg) > f$  would  identify $\sg$ and hence give a shorter description of $\sg$ than its complexity $K(\sg)$ allows.

Chaitin's Theorem gives a new and inexhaustible class of unprovable but true statements (which also lie in a very different class from those provided by G\"{o}del himself):
\begin{center}
\emph{For almost all random strings their randomness cannot be proved}.\footnote{See Raatikainen (1998) for a detailed presentation of Chaitin's incompleteness Theorem including a devastating critique of the far-reaching philosophical interpretation Chaitin himself gave of his theorem.}
\end{center}
To the extent that there are deep logico-philosophical truths, surely this is one! Compare: \begin{quote}\begin{small}
It may be taken for granted that any attempt at defining disorder in a formal way will lead to a contradiction. This does not mean that the notion of disorder is contradictory. It is so, however, as soon as I try to formalize it.\footnote{Statement by the influential German-Dutch mathematician and educator Hans Freudenthal from 1969, quoted in both van Lambalgen (1987), p.\ 8 and Terwijn (2016), p.\ 51.} 
\end{small}\end{quote}
The satisfactory definition of 1-randomness proves this wrong, but the  statement preceding it is a correct version of a similar  intuition: \emph{randomness is by its very nature elusive}.
\subsection*{Patternlessness}
This is the most direct attack on the ``paradox of probability", which states that each individual sequence has probability zero despite the huge differences in their (true or apparent) ``randomness".  We use the notation of \S\ref{PP}, so that $\mu^{\infty}$ is the probability measure on the set $\ul{2}^{\N}$ of all binary sequences induced by the 50-50 measure on the outcome space $\{0,1\}$ of a single fair coin flip (the discussion is easily adapted to other cases).

As opposed to single outcomes, a key role will be played by \emph{tests}, i.e.\ (measurable) \emph{sets} $T\subset \ul{2}^{\N}$ of outcomes for which $\mu^{\infty}(T)=0$. Such a test defines a \emph{property} that a sequence $x\in \ul{2}^{\N}$ may or may not have, namely membership of $T$. Conversely, and perhaps less circularly, one may start with a given pattern that $x$ might have, which would make it appear less random if it had it, and express this pattern as a test $T$. Conceptually, tests contain outcomes that are supposed to be ``atypical", and randomness of $x\in \ul{2}^{\N}$ will be defined by the property of \emph{not} belonging to any such test, making $x$ ``typical".  For example, the property
$\lim_{N\raw\infty}  N\inv  \sum_{n=1}^N x_n=1/2$ is ``typical" for a fair coin flip and indeed (by the strong law of large numbers) it holds with probability 1 (in that the set $E$ of all $x$ for which this limit exists and equals 1/2 has $\mu^{\infty}(E)=1$, so that its complement $T=\ul{2}^{\N}\backslash E$ has $\mu^{\infty}(T)=0$).
One has to proceed very carefully, though, since all singletons  $T=\{x\}$ for  $x\in  \ul{2}^{\N}$ are to be excluded; indeed, these \emph{all} have measure zero and yet, returning to the paradox of probability, some uncontroversially  random sequence $x$ (on whatever definition) would fail to be random by the criterion just proposed if all measure zero sets were included in the list of tests. Kolmogorov's former PhD student Martin-L\"{o}f (1966) saw his way out of
 this dilemma, combining techniques from computability/recursion theory with some ideas from the intuitionistic/constructive mathematics of the great L.E.J.\ Brouwer:
 \begin{enumerate}
\item 
 One specializes to tests of the form $T=\cap_{n\in\N} U_n$, where $U_{n+1}\subseteq U_n$ and 
 \beq
 \mu^{\infty}(U_n)\leq 2^{-n}, \label{50}
 \eeq 
 which guarantees that $\mu^{\infty}(T)=0$. 
Perhaps the simplest example is not the law of large numbers, for which see Li \& Vit\'{a}nyi (2008), \S 2.4, Example 2.4.2, but the test where $U_n$ consists of all sequences starting with $n$ zeros; clearly, $ \mu^{\infty}(U_n)= 2^{-n}$. In this case, the  test $T=\{x_{\mathrm{love}}\}$ \emph{does} consist of a singleton $x_{\mathrm{love}}=000\cdots$ (zeros only).
\item Both the sets $U_n$ and the map $n\mapsto U_n$ have to be computable in a suitable sense.\footnote{\label{Un}First, each $U_n$ has to be open in $\ul{2}^{\N}$, which means that  $U_n=\cup_{\sg\in V_n} N(\sg)$, where $V_n\subset \ul{2}^*$
and $N(\sg)=\{\sg y\mid y\in \ul{2}^{\N}\}$ consists of all sequences $x=\sg y$ that start with the given finite part $\sg$. Second, 
each $V_n$ must be countable, so that $V_n=\cup_m \sg_{(n,m)}$, where each $\sg_{(n,m)}\in \ul{2}^*$. Finally, 
there must be a single program enumerating the sets $(V_n)$, so that all in all one requires a (partial) computable function $\sg:\N\x\N\raw \ul{2}^*$ such that $U_n=\cup_m N(\sg_{(n,m)})$ and hence $T=\cap_n\cup_m N(\sg_{(n,m)})$. Such a set $T$ is called \emph{effective}, and if also \er{50} holds, then $T$ is an \emph{effective measure zero set} or \emph{Martin-L\"{o}f test}.  \label{effective}
}
\item A sequence $x\in \ul{2}^{\N}$ is  \emph{Martin-L\"{o}f random} (= \emph{typical})  if $x\notin T$ for any such test $T$.
\end{enumerate}
If $x\notin T$, one says that $x$ \emph{passes} the test $T$. Note that the computability requirement  implies that  the set of all tests $T$ satisfying these two criteria is countable, which fact by itself already shows what a huge cut in the set of all measure-zero sets has been achieved.\footnote{There is even a \emph{single} ``universal" Martin-L\"{o}f test $U$ such that $x\notin U$ iff $x$ is Martin-L\"{o}f random.}

It should be clear that this definition makes no sense if the sample space is finite, but in order to adhere to Earman's principle one could still check to what extent randomness of sequences is determined by their finite initial segments. This is indeed the case, as follows from the detailed structure of the admissible sets $U_n$ above (see footnote \ref{Un}).
\subsection*{Unpredictability}
Unpredictability is  a family resemblance, like randomness: the following definition is just one possibility, cf.\ e.g.\ footnote \ref{mrd} and eq.\ \er{CW}. Having said this:
unpredictability of a binary sequence $x = (x_1, x_2, \ldots)$ 
is formalized as the impossibility of a successful betting strategy on the successive digits of $x$. Suppose a punter with initial capital $d(0)$ starts by betting
 $b(x_1=0)\equiv b(0)$ on $x_1 = 0$ and $b(x_1 = 1)\equiv b(1)$ on $x_1 = 1$, where $b(0) + b(1) = d(0)$, with fair payoff $d(x_1) = 2b(0)$  if $x_1 = 0$ and $d(x_1) = 2b(1)$ if $x_1 = 1$. It follows that  
 \beq
 d(0) +  d(1) =2( b(0) + b(1)) = 2d(0). \label{M0}
 \eeq 
 This motivates  the concept of a \emph{martingale} as a function $d: \ul{2}^*\raw[0,\infty)$ that satisfies 
 \begin{equation}
d(\sg 0)+d(\sg 1)=2d(\sg),\label{M1}
\end{equation}
 for each $\sg\in \ul{2}^*$. Both the betting strategy itself and the payoff (for given $x\in \ul{2}^{\N}$) can be reconstructed from $d$: the bet on the $N+1$'th digit $x_{N+1}$ is given by
 \begin{align}
 b(x_{N+1}=0)&=\half d(x_{N}0);\\
 b(x_{N+1}=1)&=\half d(x_{N}1),
 \end{align} 
and after $N$ bets the punter owns $d(x_{|N})$. A martingale $d$ \emph{succeeds} on $A\subset \ul{2}^{\N}$ if
\begin{equation}
\lim\sup_{N\raw\infty} d(x_{|N})=\infty \:\:\: \mbox{ for each } x\in A, \label{punter}
\end{equation}
in which case the punter beats the casino.\footnote{In practice someone beating the casino will go home with a finite amount of money. The fact that the right-hand side of \er{punter} is infinite is the result of idealizing long strings by sequences: if the punter  has a uniform winning strategy and places infinitely many bets, he will earn an infinite amount of money. }
Our first impulse would now be to call $x\in \ul{2}^{\N}$ random if there exists no martingale that succeeds on $A=\{x\}$, but,  Ville (1939) proved:
\begin{center}
\emph{ Let $A\subset \ul{2}^{\N}$. Then $\mu^{\infty}(A)=0$ if and only if there exists a martingale that succeeds on $A$.}
\end{center}
Here $A$ should be measurable. In particular, for any sequence $x\in \ul{2}^{\N}$ there exists a martingale that succeeds on $x$, and hence no sequence $x$ would be random on the criterion that no martingale succeeds on it.  Fortunately, the previous two definitions of randomness suggest that all that is missing is a suitable notion of \emph{computability} for martingales.
This notion was provided by Schnorr (1971): all that needs to be added is that the class of martingales $d$ that succeed on $A$ be \emph{uniformly left computably enumerable}, in the sense that  firstly each real number $d(\sg)$ 
 is the limit of a computable increasing sequence of rational numbers, and secondly that there is a single program that computes $d(\sg)$ in that way. 
 
Defining $x\in \ul{2}^{\N}$ to be \emph{Schnorr random} if there exists no uniformly left computably enumerable martingale that succeeds on it,\footnote{In the literature the term `Schnorr randomness' is often used differently, namely to indicate that the martingales $d$ in the above definition are merely computable, which yields a weaker notion of randomness. Also, note that Schnorr (1971) used so-called supermartingales in his definition, for which one has $\leq$ instead of equality in \er{M1}, but martingales also work, cf.\ Downey \& Hirschfeldt (2010), Theorem 6.3.4. 
} one has the crowning theorem on 1-randomness:\footnote{The equivalence between the criteria of Martin-L\"{o}f  and of 
Kolmogorov--Chaitin was proved by Chaitin (cf.\ Calude, 2010, Theorem 6.35) and by Schnorr (1973). The equivalence between  Martin-L\"{o}f  and Schnorr is due to Schnorr (1971), Staz 5.3. 
See also
 Downey \& Hirschfeldt (2010), Theorems 6.2.3 and 6.3.4.  }
 \begin{theorem}
 A  sequence $x\in \ul{2}^{\N}$ is Kolmogorov--Chaitin random (i.e.\ incompressible) 
  iff it is Martin-L\"{o}f random (i.e.\ patternless) and iff it is Schnorr random (i.e.\ unpredictable). 
   \end{theorem}
   These equivalent conditions, then, define \emph{1-randomness} of sequences $x\in \ul{2}^{\N}$.
\subsection*{Some reflections on 1-randomness}
Any sound definition of randomness (for binary sequences) has to navigate between Scylla and Charybdis: if the definition is too weak (such as Borel normality), counterexamples will undermine it (such as Champernowne's number), but if it is too strong (such as being lawless as in Brouwer's choice sequences, cf.\ \S\ref{RFR}), it will not hold almost surely in a 50-50 Bernoulli process.
 In this respect 1-randomness does very well: see Theorem \ref{thm:3.1} and the theorem below (which  lies at the basis of Corollary \ref{CC}):
\begin{theorem}\label{PRS}
Any 1-random sequence is  Borel normal, incomputable, and contains any finite string infinitely often.\footnote{See the footnotes to Corollary \ref{CC} for the meaning of these terms, and  Calude (2010), \S 6.4 for proofs. }
\end{theorem}
In fact, since computability is sometimes used  as a mathematical metaphor for determinism, it is worth mentioning that 1-random sequences  are incomputable in a very strong sense: each 1-random sequence $x$  is \emph{immune} in the sense that neither 
the set $\{n\in\N\mid x_n=1\}$ nor its complement  $\{n\in\N\mid x_n=0\}$ is computably enumerable (c.e.), or can even contain an infinite c.e.\ subset. Thus 1-randomness is far stronger than mere incomputability (or perhaps indeterminism).
On the other hand, as already noted in \S\ref{CAC},  Chaitin's Incompleteness Theorem (see Appendix \ref{AR}) makes it impossible to prove that some given outcome sequence of a quantum-mechanical experiment with binary outcomes with 50-50 Born probabilities is 1-random \emph{even if it is} (which is worrying, because one can never be \emph{sure}, but only \emph{almost sure}, that outcomes are 1-random). At best, in a bipartite (Alice \& Bob) setting of the kind discussed in \S\ref{CAC}, one may hope for results showing that if an outcome is \emph{not} 1-random, then some kind of superluminal signaling is possible, but, as reviewed in \S\ref{CAC}, even that much has not been rigorously established so far.\footnote{See also Kamminga (2019) and references therein for a survey of the literature on this topic.} 

This suggests to try and prove randomness properties of such outcomes sequences that are on the one hand weaker than 1-randomness, like incomputability, but on the other hand are stronger, in that they hold surely. 
To this end, Abbott,  Calude, \&  Conder (2012)  take a 3-level quantum system, prepare this in some state $\psi\in\C^3$ and repeatedly (indeed, infinitely often) measure any projection $|\phv\ra\la\phv|$ for which $\sqrt{5/14}\leq |\la\phv,\psi\ra|\leq\sqrt{9/14}$; for an experimental setup realizing these measurements see Kulikov \emph{et al} (2017). In that case there is a finite set $\mathcal{P}$ of 1-dimensional projections on $\C^3$ that contains $|\phv\ra\la\phv|$ as well as $|\psi\ra\la\psi|$ and admits no coloring (Abbott,  Calude, \& Svozil, 2015).\footnote{\label{colorfn}
A \emph{coloring} would be a function $C:\mathcal{P}\raw \{0,1\}$ such that for any orthogonal set $\{e_1,e_2,e_3\}$ in  $\mathcal{P}$ with
$e_1+e_2+e_3=1_3$ (where $1_3$ is the $3\x 3$ unit matrix) there is exactly one  $e_i$ for which $C(e_i)=1$. } Some particular outcome sequence is described by a function $f:\N\raw\{0,1\}$.
The crucial assumption  Abbott \emph{et al} then make is that if $f$ is computable, then it extends to a function
$\til{f}: \N\x\mathcal{P}\raw\{0,1\}$ for which each $\til{f}_n: \mathcal{P}\raw\{0,1\}$ is a coloring of $\mathcal{P}$, where $\til{f}_n(e)=\til{f}(n,e)$. Clearly, then, on this assumption $f$ cannot be computable. This gives the desired result, but  moves the burden of proof to justifying the assumption. Phrased by the authors in terms of ``\emph{elements of physical reality}" \`{a} la EPR, their assumption in fact postulates that if $f$ is computable, then it originates in some (``morally" deterministic) non-contextual hidden-variable theory (to which their own sharpened Kochen--Specker Theorem applies).

A different line of attack  (see e.g.\ Ac\'{\i}n  \emph{et al}, 2016; Bendersky  \emph{et al}, 2016;  Herrero-Collantes \& Garcia-Escartin, 2017;  Kovalsky \emph{et al}, 2018)
assumes Hidden Locality and some amount of Freedom (cf.\ \S\ref{RQM}) to prove that 
 outcome sequences cannot originate in a hidden-variable theory  (to which Bell's Theorem applies) and hence are indeterministic. 
\section{Bell's Theorem and  Free Will Theorem}\label{FWT}
In support of the analysis of hidden variable theories in \S\ref{RQM} this appendix  reviews Bell's (1964) Theorem and the Free Will Theorem, streamlining earlier expositions  (Cator \&  Landsman, 2014;  Landsman, 2017, Chapter 6) and leaving out proofs and other adornments.\footnote{The original reference for Bell's Theorem is Bell (1964), with innumerable follow-up papers of which we especially recommend  Werner \& Wolf (2001). The Free Will theorem originates in  Heywood \& Redhead (1983), Stairs (1983),
 Brown \& Svetlichny (1990),  Clifton (1993), and Conway \& Kochen (2006, 2009). Both theorems can and have been presented and (re)interpreted in many different ways, of which we choose the one that is relevant for the general discussion on randomness in the main body of the paper. } 
 In the specific context of 't Hooft's theory (where the measurement settings are determined by the hidden state) and Bohmian mechanics (where they are not, as in the original formulation of Bell's Theorem and in most hidden variable theories) a major advantage of my approach is that both determined und undetermined stettings fall within its scope; the latter case arises from the former by adding an independence assumption.\footnote{
  This addresses a problem Bell faced even according to some of his most ardent supporters
(Norsen, 2009; Seevinck \& Uffink, 2011), namely the tension between the idea that the hidden variables (in the  causal past) should on the one hand include all ontological information relevant to the experiment, but on the other hand should leave Alice and Bob free to choose any settings they like. Whatever its ultimate fate, 't Hooft's staunch determinism has drawn attention to issues like this, as has the Free Will Theorem.}

As a warm-up I start with a version of the Kochen--Specker Theorem whose logical form is very similar to Bell's (1964) Theorem and the Free Will Theorem, as follows:
    \begin{theorem}\label{KSthm}
 Determinism, Nature, Non-contextuality, and Freedom  are contradictory.
\end{theorem}
Of course, this slightly unusual formulation hinges on the precise meaning of these terms. 
\begin{itemize}
\item 
\index{Determinism (assumption)}\hi{Determinism} is the conjunction of the following two assumptions:

1. There is a state space $X$  with associated functions $A: X\raw S$ and $L:X\raw O$,
where $S$ is  the set of all  possible \emph{measurement settings} Alice can choose from, namely 
 a suitable finite set of orthonormal bases of $\R^3$ (11 well-chosen bases will to to arrive at a contradiction),\footnote{If her setting is a basis 
 $(\vec{e}_1,\vec{e}_2,\vec{e}_3)$, Alice measures the quantities $(J_{\vec{e}_1}^2, J_{\vec{e}_2}^2, J_{\vec{e}_3}^2)$, where
 $J_{\vec{e}_1}=\la\vec{J},\vec{e}_i\ra$ is the component of the angular momentum operator $\vec{J}$ of a massive spin-1 particle in the direction $\vec{e}_i$.} and  $O$ is some set of possible  \emph{measurement outcomes}. Thus some $x\in X$ determines \emph{both} Alice's setting $a=A(x)$ \emph{and} her 
 outcome  $\al=L(x)$. 

2. There exists some set $\Lm$ and an additional function 
$H:X\raw \Lm$ such that  
\beq
L=L(A,H),
\eeq
 in the sense that
 for each $x\in X$ one has $L(x)=\hat{L}(A(x),H(x))$
  for a certain function  $\hat{L}:S \x \Lm\raw O$.  This self-explanatory assumption just states that each measurement outcome 
  $L(x)=\hat{L}(a,\lm)$ 
  is determined by the  measurement setting $a=A(x)$ and the ``hidden" variable or state $\lm=H(x)$  of the particle
 undergoing measurement.
  \item  \hi{Nature}  fixes 
 $O=\{(0,1,1), (1,0,1), (1,1,0)\}$, which is a non-probabilistic fact of \qm\ with overwhelming (though  indirect) experimental support. 
   \item \hi{Non-contextuality}
stipulates that the function $\hat{L}$ just introduced take the form
  \beq
  \hat{L}((\vec{e}_1,\vec{e}_2,\vec{e}_3),\lm)=(\til{L}(\vec{e}_1,\lm), \til{L}(\vec{e}_2,\lm), \til{L}(\vec{e}_3,\lm)), \label{hatL3}
  \eeq
  for a single function $\til{L}:S^2\x \Lm\raw\{0,1\}$ that also satisfies 
$\til{L}(-\vec{e},\lm)=\til{L}(\vec{e},\lm)$.\footnote{Here $S^2=\{(x,y,z)\in\R^3\mid x^2+y^2+z^2=1\}$ is the 2-sphere, seen as the space of unit vectors in $\R^3$.
Eq.\ \er{hatL3} means that the outcome of Alice's measurement 
of $J_{\vec{e}_i}^2$ is independent of the ``context" $(J_{\vec{e}_1}^2, J_{\vec{e}_2}^2, J_{\vec{e}_3}^2)$; she might as well measure $J_{\vec{e}_i}^2$ by itself. The last equation is trivial, since $(J_{-\vec{e}_i})^2=(J_{\vec{e}_i})^2$.
}
  \item \hi{Freedom} finally states that the following function is surjective:
\begin{align}
A\x H:X\raw S\x \Lm; && x\mapsto (A(x),H(x)).
\end{align}
 In other words, 
 for each $(a,\lm)\in S\x\Lm$ there is an $x\in X$  for which $A(x)=a$ and $H(x)=\lm$. This makes   $A$ and $H$ are ``independent"\
 (or: $a$ and $\lm$ are free variables).
  \end{itemize}
See Landsman (2017), \S6.2 for a proof of the Kochen--Specker Theorem in this language.\footnote{
The assumptions imply the existence of a coloring $C_{\lm}: \mathcal{P}\raw\{0,1\}$ (cf.\ footnote \ref{colorfn}), where 
$\CP\subset S^2$ consist of all unit vectors contained in all bases in $S$, and $\lm$ ``goes along for a free ride". Indeed, one finds
$C_{\lm}(\vec{e})=\til{L}(\vec{e},\lm)$. The point, then, is that on a suitable choice of the set $S$ such a coloring cannot exist. 
}
\smallskip

\noindent Bell's (1964) Theorem and the Free Will Theorem both take a similar generic form, namely:
    \begin{theorem}\label{FWTthm}
Determinism, Nature, (Hidden) Locality, and Freedom,   are contradictory.
\end{theorem}
Once again, I have to explain what these terms exactly mean in the given context. 
\begin{itemize}
\item \hi{Determinism} is a straightforward adaptation of the above meaning to the bipartite ``Alice and Bob" setting. Thus we have a state space $X$  with associated functions 
\begin{align}
A: X\raw S; & &
B: X\raw S; &&
 L:X\raw O & &
 R: X\raw O, \label{C4}
\end{align}
where the set $S$ of all possible measurement settings Alice and Bob can each choose from differs a bit between the two theorems:
for the Free Will Theorem it is the same as for the Kochen--Specker Theorem above, as is the set  $O$ of possible measurement outcomes, 
whereas for Bell's Theorem (in which Alice and Bob each measure a 2-level system), $S$ is some finite set of angles (three is enough), and 
 $O=\{0,1\}$.
 \begin{itemize}
\item In the Free Will case, these functions and the state $x\in X$ determine both the settings $a=A(x)$ and $b=B(x)$ of a measurement  and its outcomes  $\al=L(x)$ and $\beta=R(x)$ for Alice on the \emph{L}eft and for Bob on the \emph{R}ight, respectively.
\item  Although this is also true in the Bell case,  his theorem relies on measurement statistics (as opposed to individual outcomes), so that one in addition assumes a probability measure $\mu$ on $X$ (sampled by repeated measurements, see \S\S\ref{PP}--\ref{CAC}).\footnote{The existence of $\mu$ is of course predicated on $X$ being a measure space with corresponding $\sg$-algebra of measurable subsets, with respect to which all functions in \er{C4} and below are measurable.}
\end{itemize}
Furthermore, there exists some set $\Lm$ and some  function  $H:X\raw \Lm$ such that
 \begin{align}
 L=L(A,B,H); &&
 R=R(A,B,H),
\end{align}
 in the sense that for each $x\in X$ one has  functional relationships
 \begin{align}
 L(x)=\hat{L}(A(x),B(x),H(x)); &&
 R(x)=\hat{R}(A(x),B(x),H(x)),\label{GhatG}
\end{align}
  for certain functions
  $\hat{L}:S \x S\x \Lm\raw O$ and  $\hat{R}:S \x S\x \Lm\raw O$. 
  \item \hi{Nature}  reflects elementary \qm\ of correlated 2-level and 3-level quantum systems for the Bell and the Free Will cases, respectively, as follows:\footnote{In Bell's Theorem quantum theory can be replaced by experimental support (Hensen  \emph{et al}, 2015). }
   \begin{itemize}
\item  In the \emph{Free Will Theorem},  $O=\{(0,1,1), (1,0,1), (1,1,0)\}$ is the same as for the Kochen--Specker Theorem. In addition  \emph{perfect correlation} obtains: if $a=(\vec{e}_1,\vec{e}_2,\vec{e}_3)$ is Alice's orthonormal basis
and  $b=(\vec{f}_1,\vec{f}_2,\vec{f}_3)$ is Bob's, one has
 \begin{equation}
\vec{e}_i=\vec{f}_j\: \Raw\: \hat{L}_i(a,b,z)=\hat{R}_j(a,b,z), \label{ienj}
\end{equation}
where $\hat{L}_i, \hat{R}_j: S \x S\x \Lm\raw \{0,1\}$ are the components of $\hat{L}$ and $\hat{R}$, respectively.  Finally,\footnote{As in Kochen--Specker, this is because Alice \& Bob measure \emph{squares} of (spin-1) angular momenta.} if $(a',b'$) differs from $(a,b)$ by changing the sign of any  basis vector,
 \begin{align}
\hat{L}(a',b',\lm)=\hat{L}(a,b,\lm); && 
\hat{R}(a',b',\lm)=\hat{R}(a,b,\lm).
\end{align}
\item In \emph{Bell's Theorem}, $O=\{0,1\}$, and the statistics for the experiment is reproduced as conditional joint probabilities given by the measure $\mu$ through
\begin{equation}
P(L\neq R|A=a,B=b)=\sin^2(a-b).\label{uitkomstAspect2} 
\end{equation}
\end{itemize}
\item \hi{(Hidden)  Locality}, which replaces Non-contextuality in Theorem \ref{KSthm}, means that 
\begin{align}
L(A,B,H)=L(A,H); && R(A,B,H)=R(B,H).
\end{align} In words: Alice's outcome \emph{given $\lm$} does not depend on Bob's setting, and \emph{vice versa}. 
\item  \hi{Freedom} is an independence assumption that looks different for both theorems:
\begin{itemize}
\item In the \emph{Free Will Theorem} it means  that  each $(a,b,\lm)\in S\x S\x \Lm$ is possible in that there is an $x\in X$  for which $A(x)=a$, $B(x)=b$, and $H(x)=\lm$.
\item In  \emph{Bell's Theorem},  $(A,B,H)$ are \emph{probabilistically independent} relative to $\mu$.\footnote{This implies  that each pair $(A,B)$, $(A,H)$, and $(B,H)$ is independent. See also Esfeld (2015).}
\end{itemize}
  \end{itemize}
This concludes the joint statement of the Free Will Theorem and Bell's (1964) Theorem in the somewhat unusual form we need for the main text. The former is proved by reduction to the Kochen--Specker Theorem, whilst the latter follows by reduction to the usual version of Bell's Theorem via the Freedom assumption; see Landsman (1917), Chapter 6 for details. 

For our purposes these theorems are equivalent, despite subtle differences in their assumptions. Bell's Theorem is much more robust in that it does not rely on perfect correlations (which are hard to realize experimentally), and in addition it requires almost no input from quantum theory.
On the other hand,  Bell's Theorem uses  probability theory in a highly nontrivial way: like the hidden variable theories it is supposed to exclude it relies on the possibility of fair sampling of the probability measure $\mu$. The 
  factorization condition defining probabilistic independence passes this requirement of fair sampling on to both the hidden variable and the settings, which brings us back to the discussion in \S\ref{CAC}. 
 \smallskip
 
 \noindent All trusting \emph{Nature},  different parties may now be identified by the assumption they drop:
 \begin{itemize}
\item Following Copenhagen \qm, most physicists reject \emph{Determinism};
\item Bohmians reject  \emph{(Hidden) Locality};
\item 't Hooft rejects  \emph{Freedom}. 
\end{itemize}
However, as we argued in \S\ref{CAC}, even the latter two camps do not really have a deterministic theory underneath \qm\ because of their need to randomly sample the probability measure they must use to recover the predictions of \qm. 

  \newpage
\addcontentsline{toc}{section}{References}
\begin{small}

\end{small}
\end{document}